\documentclass{emulateapj}
\usepackage{apjfonts}
\usepackage{amsmath}

\usepackage{graphics}

\newcommand{\bd}{\begin{displaymath}}
\newcommand{\ed}{\end{displaymath}}

\slugcomment{Received 2011 March 8; accepted 2011 June 12}
\shorttitle{The large scale magnetic fields of ADAFs}

\begin{document}

\title{The large scale magnetic fields of advection dominated accretion flows }

%\author{Xinwu Cao\altaffilmark{1}}
\author{Xinwu Cao}
\affil{Key Laboratory for Research in Galaxies and Cosmology,
Shanghai Astronomical Observatory, Chinese Academy of Sciences, \\80
Nandan Road, Shanghai, 200030, China; cxw@shao.ac.cn}

\begin{abstract}
We calculate the advection/diffusion of the large-scale magnetic
field threading an advection dominated accretion flow (ADAF), and
find that the magnetic field can be dragged inward by the accretion
flow efficiently, if the magnetic Prandtl number ${\cal P}_{\rm
m}=\eta/\nu\sim 1$. This is due to the large radial velocity of the
ADAF. It is found that the magnetic pressure can be as high as $\sim
50\%$ of the gas pressure in the inner region of the ADAF close to
the black hole horizon, even if the external imposed homogeneous
vertical field strength is $\la 5\%$ of the gas pressure at the
outer radius of the ADAF, which is caused by the gas in the ADAF
plunging rapidly to the black hole within the marginal stable
circular orbit. In the inner region of the ADAF, the accretion flow
is significantly pressured in the vertical direction by the magnetic
fields, and therefore its gas pressure can be two orders of
magnitude higher than that in the ADAF without magnetic fields. This
means that the magnetic field strength near the black hole is
underestimated by assuming equipartition between magnetic and gas
pressure with the conventional ADAF model. Our results show that the
magnetic field strength of the flow near the black hole horizon can
be more than one order of magnitude higher than that in the ADAF at
$\sim 3R_{\rm g}$ ($R_{\rm g}=2GM/c^2$), which implies the
Blandford-Znajek mechanism could be more important than the
Blandford-Payne mechanism for ADAFs. We find that the accretion flow
is decelerated near the black hole by the magnetic field when the
external imposed field is strong enough or the gas pressure of the
flow is low at the outer radius, or both. This corresponds to a
critical accretion rate, below which the accretion flow will be
arrested by the magnetic field near the black hole for a given
external imposed field. In this case, the gas may accrete as
magnetically confined blobs diffusing through field lines in the
region very close to the black hole horizon, which is similar to
those in compact stars. Our calculations are also valid for the case
that the inner ADAF connects to the outer cold thin disk at a
certain radius. In this case, the advection of the external fields
is quite inefficient in the outer thin disk due to its low radial
velocity, and the field lines thread the disk almost vertically,
while these field lines can be efficiently dragged inward by the
radial motion of the inner ADAF.
\end{abstract}

\keywords{accretion, accretion disks, galaxies: jets, magnetic
fields, galaxies: active}

%\altaffiltext{1}{Key Laboratory for Research in Galaxies and
%Cosmology, Shanghai Astronomical Observatory, Chinese Academy of
%Sciences, 80 Nandan Road, Shanghai, 200030, China; cxw@shao.ac.cn}

\section{Introduction}

The winds driven from the accretion disk through the magnetic field
lines threading the disk have been considered as promising
explanations for jets/outflows observed in different types of the
sources, such as, active galactic nuclei (AGNs), X-ray binaries, and
young stellar objects \citep*[see reviews
in][]{1996epbs.conf..249S,2000prpl.conf..759K,2007prpl.conf..277P,2010LNP...794..233S}.
In this model, the ordered magnetic field co-rotates with the gases
in the disk, and the jets are powered by the gravitation energy
released by accretion of the gases through the ordered field
threading the disk \citep{1982MNRAS.199..883B}. The ordered
large-scale magnetic field threading the disk is the crucial
ingredient in this model. In most of the previous works, the
strength of the magnetic field is simply assumed to scale with the
gas/radiation pressure of the accretion disk
\citep*[e.g.,][]{1996MNRAS.283..854M,1997MNRAS.292..887G,1999ApJ...512..100L,1999ApJ...523L...7A,2007MNRAS.377.1652N,2008ApJ...687..156W,2009ApJ...698..594M}.
It was suggested that, the external generated large-scale poloidal
field (e.g., the field originates from the interstellar medium)
would be dragged inward by the accretion plasma, while the field
will diffuse outward at the same time
\citep*[][]{1974Ap&SS..28...45B,1976Ap&SS..42..401B,1989ASSL..156...99V,1994MNRAS.267..235L,2001ApJ...553..158O}.
In this case, the final steady magnetic field configuration can be
derived, in which the inward advection of the field lines is
balanced by the outward movement of field lines due to magnetic
diffusion \citep{1994MNRAS.267..235L}. This means that the magnetic
field configuration is dominantly determined by the radial velocity
of the disk and magnetic diffusivity. As the radial velocity of the
accretion disk is roughly proportional to the kinematic viscosity
$\nu$, the magnetic field configuration is sensitive to the magnetic
Prandtl number ${\cal P}_{\rm m}=\eta/\nu$, where $\eta$ is magnetic
diffusivity.

\citet{parker1979} argued that $\nu\sim\eta\sim lv_{\rm t}$ ($l$ is
the largest eddy size, and $v_{\rm t}$ is turnover velocity), and
${\cal P}_{\rm m}=\eta/\nu\sim 1$, is expected in isotropic
turbulence. This issue was explored by some different authors with
numerical simulations
\citep*[e.g.,][]{2003A&A...411..321Y,2009A&A...504..309L,2009A&A...507...19F,2009ApJ...697.1901G},
which all suggest that the magnetic Prandtl number should be around
unity.  A suitable magnetic field configuration is crucial for
launching a jet from the accretion disk. More specifically, the
angle of field lines inclined to the midplane of the disk being less
than $\sim 60^\circ$ is required for launching jets from a Keplerian
cold disk \citep*{1982MNRAS.199..883B,1994A&A...287...80C}. This
critical angle could be larger than $60^\circ$ for the accretion
disk surrounding a rapidly spinning black hole
\citep{1997MNRAS.291..145C}, which indicates that the spin of black
hole may help launching jets centrifugally by cold magnetized disks
\citep{1997MNRAS.291..145C,2010A&A...517A..18S}.
\citet{1994MNRAS.267..235L} explored the final steady magnetic field
configuration with the balance between advection and diffusion of
the large-scale magnetic field, and found that significant inward
dragging of fields occurs only if ${\cal P}_{\rm m}\la H/R$ is
satisfied ($H$ is the scale-height of the disk at radius $R$). This
means that the dragging of the external fields is always unimportant
for thin disks, as ${\cal P}_{\rm m}\sim 1$ is suggested, however,
the advection of magnetic fields may be efficient for a
geometrically thick accretion disk with $H\sim R$
\citep*[e.g.,][]{2011arXiv1101.2292F}. An alternative model was
suggested by \citet{2005ApJ...629..960S} for advection of the
external field in the disk, in which turbulent diffusion is reduced
by bundles of the large-scale magnetic field
\citep{2001MNRAS.323..587S}. \citet{2009ApJ...701..885L} suggested
that the field can be efficiently advected inward based on the
assumption of the surface layer of the accretion disk to be
nonturbulent. The general relativistic magnetohydrodynamic (GRMHD)
simulation of an accretion torus embedded in a large-scale magnetic
field showed that the mass is accreted mainly within the accretion
disk, and the magnetic field flux is carried by the motions in the
low-density corona \citep{2009ApJ...707..428B}.

Low mass accretion rate $\dot{m}$ may lead to the accretion flows to
be advection-dominated
\citep{1994ApJ...428L..13N,1995ApJ...452..710N}. Advection dominated
accretion flows (ADAFs) are suggested to be present in
low-luminosity active galactic nuclei (AGNs) \citep*[see][for a
review and references therein]{2002luml.conf..405N}, FR I radio
galaxies \citep*[e.g.,][]{2001A&A...379L...1G,2004MNRAS.349.1419C},
or BL Lac objects \citep*[e.g.][]{2003ApJ...599..147C}. The ADAF
model can successfully explain most observational features of
low-luminosity AGNs and black hole X-ray binaries in the low/hard
state
\citep*[e.g.,][]{1996ApJ...462..142L,1999ApJ...516..177G,1999ApJ...525L..89Q,2009RAA.....9..401X}.
ADAFs are hot and geometrically thick, which have relatively higher
radial velocities than thin accretion disks
\citep{1994ApJ...428L..13N,1995ApJ...452..710N}. This implies that
the advection of the external fields in ADAFs may be more
efficiently than that in thin disks.

In this work, we explore the advection/diffusion of the large-scale
magnetic fields threading an ADAF, in which the compression of the
accretion flow in the vertical direction by the magnetic field is
properly considered.

\section{Model}

\subsection{The structure of the ADAF}

The dynamics of a steady ADAF with large-scale magnetic fields are
described by a set of differential equations, namely, the continuity
equation, the radial and azimuthal momentum equations, and the
energy equation. {In this work, we use cylindrical coordinates
$(R,\phi,z)$. }

The continuity equation is
\begin{equation}
{\frac {d}{dR}}(\rho RHv_R)=0, \label{continuity_1}
\end{equation}
where mass loss rate in the winds from the ADAF is neglected.

The radial momentum equation is
\begin{equation}
v_R{\frac {dv_R}{dR}}=-(\Omega_{\rm K}^2-\Omega^2)R-{\frac
1{\rho}}{\frac {d}{dR}}(\rho c_{\rm s}^2)+{\frac {B_R^{\rm
S}B_z}{2\pi \Sigma}}-{\frac {B_z H}{2\pi\Sigma}}{\frac {\partial
B_z}{\partial R}}, \label{radial_1}
\end{equation}
where $B_R^{\rm S}$ is the radial component of the large-scale
magnetic fields at the disk surface, and the pseudo-Newtonian
potential is adopted to simulate the gravity of a non-rotating black
hole \citep{1980A&A....88...23P}. {The term ${B_R^{\rm S}B_z}/{2\pi
\Sigma}$ is the radial component of the force caused by curvature of
the field lines, which is derived from $\int_{-H}^{H}
(B_{z}/4\pi\Sigma)\partial B_{R}/\partial z dz$ by approximating
$\partial B_{R}/\partial z\simeq B_{R}^{\rm S}/H$.} The Keplerian
angular velocity is given by
\begin{equation}
\Omega_{\rm K}^2(R)={\frac {GM}{(R-R_{\rm g})^2 R}}, \label{omega_k}
\end{equation}
where $R_{\rm g}=2GM/c^2$.

The radial magnetic fields will be sheared into azimuthal fields by
the differential rotation of the flow, which leads to
magnetorotational instability (MRI) and the MRI-driven turbulence
\citep{1991ApJ...376..214B,1998RvMP...70....1B}. For simplicity, we
assume that the conventional $\alpha$-viscosity can describe the
angular momentum transportation in the accretion flow caused by
MRI-driven turbulence. The angular momentum equation is
\begin{equation}
{\frac {d\Omega}{dR}}={\frac {v_R(j-j_{\rm in})}{\nu R^2}},
\label{azimuthal_1}
\end{equation}
where $j$ is the specific angular momentum of the flow, $j_{\rm in}$
is the specific angular momentum of the gas swallowed by the black
hole, and the $\alpha$-viscosity,
\begin{equation}
\nu=\alpha c_{\rm s}H, \label{viscosity}
\end{equation}
is adopted. The angular momentum equation (\ref{azimuthal_1})
reduces to an algebraic equation,
\begin{equation}
j={\frac {v_R{R}j_{\rm in}}{\alpha c_{\rm s}H+v_{R}R}},
\label{azimuthal_2}
\end{equation}
by assuming $d\Omega/dR\simeq -\Omega/R$.

The energy equation is
\begin{equation}
\Sigma v_R T{\frac {ds}{dR}}=Q^{+}-Q^{-}, \label{energy_1}
\end{equation}
where $T$ is the temperature and $s$ is the entropy of the gas in
the ADAF. Equation (\ref{energy_1}) can be re-written as
\begin{equation}
{\frac {2}{(\gamma-1)c_{\rm s}}} {\frac {dc_{\rm s}}{dR}}={\frac
{1}{\rho}}{\frac {d\rho}{dR}}+{\frac {fv_R}{\alpha c_{\rm
s}^3R^2H}}(j-j_{\rm in})^2, \label{energy_2}
\end{equation}
%\begin{equation}
%{\frac {\gamma+1}{(\gamma-1)}} {\frac {1}{ c_{\rm s}}}{\frac
%{dc_{\rm s}}{dR}}=-{\frac {1}{R}}-{\frac {1}{v_r}}{\frac
%{dv_r}{dR}}+{\frac {1}{{\Omega}_{\rm K}}}{\frac {d\Omega_{\rm
%K}}{dR}}+{\frac {fv_r}{\alpha c_{\rm s}^3R^2H}}(j-j_{\rm in})^2,
%\label{energy_2}
%\end{equation}
where $\gamma$ is the ratio of specific heats, and the parameter $f$
describes the fraction of the dissipated energy advected in the
flow. Substituting the continuity equation (\ref{continuity_1}) into
Equation (\ref{energy_2}), we have
\begin{equation}
{\frac {2}{(\gamma-1)c_{\rm s}}} {\frac {dc_{\rm s}}{dR}}=-{\frac
{1}{R}}-{\frac {1}{H}}{\frac {dH}{dR}}-{\frac {1}{v_R}}{\frac
{dv_R}{dR}}+{\frac {fv_R}{\alpha c_{\rm s}^3R^2H}}(j-j_{\rm in})^2.
\label{energy_3}
\end{equation}
The vertical structure of the accretion flow is significantly
altered in the presence of large-scale magnetic fields. In this
work, we calculate the vertical structure of the ADAF following the
approach given in \citet{2002A&A...385..289C}. Assuming the
accretion flow to be isothermal in the vertical direction, we can
calculate the vertical structure of the ADAF with
\begin{equation}
c_{\rm s}^2{\frac {d\rho(z)}{dz}}=-\rho(z)\Omega_{\rm K}^2z-{\frac
{B_{R}}{4\pi}}{\frac {\partial B_R}{\partial z}}+{\frac
{B_{R}}{4\pi}}{\frac {\partial B_z}{\partial R}}, \label{vertical_1}
\end{equation}
provided the shape of the magnetic field lines in the accretion flow
is known. An additional term containing $\partial B_z/\partial R$,
{which is the vertical component of the force caused by curvature of
the field lines,} is included, because $H\sim R$ holds at least in
the outer region of ADAFs. {For geometrically thin accretion disks
($H\ll R$), $\partial B_{R}/\partial z\gg
\partial B_z/\partial R$, and the term containing $\partial
B_z/\partial R$ is therefore neglected \citep{2002A&A...385..289C}}

In principle, the field line shape is computable by solving the
radial and vertical momentum equations with suitable boundary
conditions \citep*[see][for the detailed
discussion]{2002A&A...385..289C}. For the isothermal case, an
approximate analytical expression is proposed for the shape of the
field lines in the flow:
\begin{equation}
R-R_{\rm i}={\frac {H}{\kappa_0\eta_{\rm i}^2}}(1-\eta_{\rm
i}^2+\eta_{\rm i}^2z^2H^{-2})^{1/2}-{\frac {H}{\kappa_0\eta_{\rm
i}^2}}(1-\eta_{\rm i}^2)^{1/2}, \label{b_shape_1}
\end{equation}
where $H$ is the scale height of the disk, $\eta_{\rm i}=\tanh(1)$,
and the inclination of the field line $\kappa_0=B_z/B_R^{\rm S}$ at
the disk surface $z=H$ \citep{2002A&A...385..289C}. This expression
can reproduce the basic features of the Kippenhahn-Schl\"{u}ter
model \citep{1957ZA.....43...36K} well either for weak or strong
field cases. The vertical distribution of the accretion disk with
magnetic fields is not exactly the Gaussian distribution of an
isothermal disk. In this case, we define the disk scale height $H$
as $\rho(H)=\rho(0)\exp(-1/2)$, the disk scale height $H$ can then
be evaluated numerically with the given magnetic field line shape
(\ref{b_shape_1}). As done by \citet{2002A&A...385..289C}, we use a
fitting formula to calculate the scale height of the disk in the
rest of this work,
\begin{equation}
{\frac {H}{R}}={\frac {1}{2}}\left ({\frac {4c_{\rm
s}^2}{R^2\Omega_{\rm K}^2}}+{f_1}^2\right)^{1/2}-{\frac {1}{2}}f_1,
\label{h_1}
\end{equation}
where
\begin{equation}
f_1={\frac {1}{2(1-{\rm e}^{-1/2})\kappa_0}}\left({\frac
{B_z^2}{4\pi\rho{R}H\Omega_{\rm K}^2\kappa_0}}+{\frac {\xi_{\rm
bz}B_z^2}{4\pi\rho{R^2}\Omega_{\rm K}^2}}\right),\label{f_1}
\end{equation}
and $\xi_{\rm bz}$ is defined as
\begin{equation}
{\frac {\partial B_z}{\partial R}}=-\xi_{\rm bz}(R){\frac {B_z}{R}},
\label{xi_bz}
\end{equation}
which can be calculated with the balance between the advection and
diffusion of the fields in the accretion flow (see Section 2.2). We
find that this formula can reproduce the numerical results quite
well. We note that equation (\ref{h_1}) reduces to $H=c_{\rm
s}/\Omega_{\rm K}$ in the absence of magnetic field.

We define dimensionless quantities by
\begin{displaymath}
r={\frac {R}{R_{\rm g}}},~~~~~\tilde{H}={\frac
{H}{R}},~~~~~\tilde{v}_R={\frac {v_R}{c}},~~~~~\tilde{c}_{\rm
s}={\frac {c_{\rm s}}{c}},~~~~~
\end{displaymath}
\begin{displaymath}
\tilde{j}={\frac {j}{R_{\rm g}c}},~~~~~\tilde{\Omega}={\frac
{\Omega}{R_{\rm g}^3c}},~~~~~\tilde{\Sigma}={\frac
{\Sigma}{\Sigma(R_{\rm out})}},~~~~~
\end{displaymath}
\begin{equation}
\tilde{B}_z={\frac {B_z}{B_0}},~~~~~\tilde{B}_R={\frac
{B_R}{B_0}},~~~~~\beta_0=p(R_{\rm out})/{\frac {B_{0}^2}{8\pi}},
\end{equation}
where $B_0$ is the strength of the putative external imposed
homogeneous vertical magnetic fields, and $R_{\rm out}$ is the outer
radius of the accretion disk.

%We re-write the equations in dimensionless form as follows.

{ Differentiate Equation (\ref{h_1}), we obtain
\begin{displaymath}
{\frac {1}{H}}{\frac {dH}{dR}}={\frac {2c_{\rm
s}}{f_{3}f_{4}H\Omega_{\rm K}^2(2H+f_{1}R)}}{\frac {dc_{\rm
s}}{dR}}-{\frac {Rf_{1}}{f_{3}f_{4}v_{R}(2H+f_{1}R)}}{\frac
{dv_R}{dR}}
\end{displaymath}
\begin{displaymath}
-{\frac {1}{f_{4}}}\left[{\frac {2c_{\rm s}^2}{f_{3}\Omega_{\rm
K}^2H(2H+f_{1}R)}}-{\frac {2f_{1}R}{f_{3}(2H+f_{1}R)}} \right]{\frac
{1}{\Omega_{\rm K}}}{\frac {d\Omega_{\rm K}}{dR}}
\end{displaymath}
\begin{equation}
-{\frac {2c_{\rm s}^2}{f_{3}f_{4}RH\Omega_{\rm
K}^{2}(2H+f_{1}R)}}-{\frac
{(f_{2}R-f_{3}H)(2H+f_{1}R)-f_{1}RH}{f_{3}f_{4}RH(2H+f_{1}R)}},
\label{dhdr_1}
\end{equation}
where
\begin{displaymath}
f_{2}=-{\frac {H}{8\pi(1-{\rm e}^{-1/2})(2H+f_{1}R)\rho R\Omega_{\rm
K}^2}}~~~~~~~~~~~~~~
\end{displaymath}
\begin{equation}
\times\left({\frac {2\xi_{\rm br}{B_{R}^{\rm
S}}B_z}{H\kappa_0}}+{\frac {\xi_{\rm bz}^2B_{R}^{\rm
S}B_{z}}{R}}+{\frac {\xi_{\rm bz}\xi_{\rm br}B_{R}^{\rm
S}B_{z}}{R}}\right),\label{f_2}
\end{equation}
\begin{equation}
f_{3}=1-{\frac {{B_{R}^{\rm S}}B_z}{8\pi(1-{\rm
e}^{-1/2})(2H+f_{1}R)\rho H\Omega_{\rm K}^2\kappa_0}},\label{f_3}
\end{equation}
\begin{equation}
f_{4}=1+{\frac {f_{1}R}{f_{3}(2H+f_{1}R)}},\label{f_4}
\end{equation}
and $\xi_{\rm br}$ is defined as
\begin{equation}
{\frac {\partial B_{R}^{\rm S}}{\partial R}}=-\xi_{\rm br}(R){\frac
{B_{R}^{\rm S}}{R}}. \label{xi_br}
\end{equation}
Substitute Equation (\ref{dhdr_1}) into the energy equation
(\ref{energy_3}), we finally obtain
\begin{displaymath}
{\frac {1}{c_{\rm s}}}{\frac {dc_{\rm s}}{dR}}={\frac
{1}{f_{4}f_{5}}}\left[{\frac {2c_{\rm s}^2}{f_{3}H\Omega_{\rm
K}^2(2H+f_{1}R)}}-{\frac {2f_{1}R}{f_{3}(2H+f_{1}R)}} \right]{\frac
{1}{\Omega_{\rm K}}}{\frac {d\Omega_{\rm K}}{dR}}
\end{displaymath}
\begin{displaymath}
-\left[1-{\frac {f_{1}R}{f_{3}f_{4}(2H+f_{1}R)}}\right]{\frac {1}
{f_{5}v_{R}}}{\frac {dv_R}{dR}}+{\frac {fv_{R}}{f_{5}\alpha c_{\rm
s}^3R^2H}}(j-j_{\rm in})^2
\end{displaymath}
\begin{displaymath}
+{\frac
{(f_{2}R-f_{3}H-f_{3}f_{4}H)(2H+f_{1}R)-f_{1}RH}{f_{3}f_{4}f_{5}RH(2H+f_{1}R)}}~~~~~~~~~~~~~~~~~~~~~
\end{displaymath}
\begin{equation}
+{\frac {2c_{\rm s}^2}{f_{3}f_{4}f_{5}RH\Omega_{\rm
K}^2(2H+f_1R)}},~~~~~~~~~~~~~~~~~~~~~~~~~~~\label{energy_4}
\end{equation}
where
\begin{equation}
f_{5}={\frac {2}{\gamma-1}}+{\frac {2c_{\rm
s}^2}{f_{3}f_{4}H\Omega_{\rm K}^2(2H+f_{1}R)}},\label{f_5}
\end{equation}
In the same way, we can re-write the radial momentum equation as
\begin{displaymath}
{\frac {v_{R}^2-v_{R,\rm s}^2}{v_R}}{\frac {dv_R}{dR}}=-R\Omega_{\rm
K}^2+{\frac {j^2}{R^3}} +{\frac {c_{\rm s}^2}{R}}+{\frac {B_R^{\rm
S}B_z}{2\pi \Sigma}}+{\frac {\xi_{\rm bz}B_z^2 H}{2\pi\Sigma
R}}~~~~~~~~~~~~
\end{displaymath}
\begin{displaymath}
{\frac {c_{\rm s}^2(f_6-f_5)}{f_4f_5}}\left[{\frac {2c_{\rm
s}^2}{f_{3}H\Omega_{\rm K}^2(2H+f_{1}R)}}-{\frac
{2f_{1}R}{f_{3}(2H+f_{1}R)}} \right]{\frac {1}{\Omega_{\rm
K}}}{\frac {d\Omega_{\rm K}}{dR}}
\end{displaymath}
\begin{displaymath}
-{\frac {[(f_{2}R-f_{3}H)(2H+f_{1}R)-f_{1}RH]c_{\rm
s}^2}{f_{3}f_{4}RH(2H+f_{1}R)}}-{\frac {2c_{\rm
s}^4}{f_3f_4RH\Omega_{\rm K}^2(2H+f_1R)}}
\end{displaymath}
\begin{displaymath}
+{\frac {2f_6c_{\rm s}^4}{f_3f_4f_5RH\Omega_{\rm
K}^2(2H+f_1R)}}+{\frac {f_6fv_{R}}{f_{5}\alpha c_{\rm
s}R^2H}}(j-j_{\rm in})^2
\end{displaymath}
\begin{equation}
+{\frac {f_6c_{\rm
s}^2[(f_{2}R-f_{3}H-f_{3}f_{4}H)(2H+f_{1}R)-f_{1}RH]}{f_{3}f_{4}f_{5}RH(2H+f_{1}R)}},\label{radial_2}
\end{equation}
where
\begin{equation}
v_{R,\rm s}=\left\{1-{\frac {f_1R}{f_3f_4(2H+f_1R)}}-{\frac
{f_6}{f_5}}\left[1-{\frac {f_1R}{f_3f_4(2H+f_1R)}}  \right]
\right\}^{1/2}c_{\rm s}, \label{vrs_1}
\end{equation}
and
\begin{equation}
f_6={\frac {2c_{\rm s}^2}{f_3f_4H\Omega_{\rm
K}^2(2H+f_1R)}}-2.\label{f_6}
\end{equation}
It is found that Equation (\ref{vrs_1}) reduces to
\begin{equation}
v_{R,\rm s}=\left({\frac {2\gamma}{\gamma+1}}\right)^{1/2}c_{\rm
s},\label{vrs_2}
\end{equation}
in the absence of magnetic field. We re-write the equations in
dimensionless form as follows.}

{The radial momentum equation is
\begin{displaymath}
{\frac {1}{\tilde{v}_R}}{\frac {d\tilde{v}_R}{dr}}={\frac
{1}{\tilde{v}_{R}^2-\tilde{v}_{R,\rm
s}^2}}\left\{-r\tilde{\Omega}_{\rm K}^2+{\frac {\tilde{j}^2}{r^3}}
+{\frac {\tilde{c}_{\rm s}^2}{r}}
 +{\frac {2r\tilde{c}_{\rm
s,out}^2\tilde{v}_R\tilde{B}_{R}^{\rm S}\tilde{B}_z}{r_{\rm
out}^2\tilde{H}_{\rm out}\tilde{v}_{R,\rm out}\beta_0}}\right.
\end{displaymath}
\begin{displaymath}
+{\frac {2r\tilde{H}\tilde{c}_{\rm s,out}^2\tilde{v}_R\xi_{\rm
bz}\tilde{B}_z^2}{r_{\rm out}^2\tilde{H}_{\rm out}\tilde{v}_{R,\rm
out}\beta_0}}-{\frac
{[(f_{2}-f_{3}\tilde{H})(2\tilde{H}+f_{1})-f_{1}\tilde{H}]\tilde{c}_{\rm
s}^2}{f_{3}f_{4}\tilde{H}(2\tilde{H}+f_{1})}}
\end{displaymath}
\begin{displaymath}
{\frac {(f_6-f_5)\tilde{c}_{\rm s}^2}{f_4f_5}}\left[{\frac
{2\tilde{c}_{\rm s}^2}{f_{3}r^2\tilde{H}\tilde{\Omega}_{\rm
K}^2(2\tilde{H}+f_{1})}}-{\frac {2f_{1}}{f_{3}(2\tilde{H}+f_{1})}}
\right]{\frac {1}{\tilde{\Omega}_{\rm K}}}{\frac
{d\tilde{\Omega}_{\rm K}}{dr}}
\end{displaymath}
\begin{displaymath}
-{\frac {2\tilde{c}_{\rm s}^4}{f_3f_4r^3\tilde{H}\tilde{\Omega}_{\rm
K}^2(2\tilde{H}+f_1)}}+{\frac {2f_6\tilde{c}_{\rm
s}^4}{f_3f_4f_5r^3\tilde{H}\tilde{\Omega}_{\rm
K}^2(2\tilde{H}+f_1)}}
\end{displaymath}
\begin{displaymath}
+{\frac {f_6\tilde{c}_{\rm
s}^2[(f_{2}-f_{3}\tilde{H}-f_{3}f_{4}\tilde{H})(2\tilde{H}+f_{1})-f_{1}\tilde{H}]}{f_{3}f_{4}f_{5}r\tilde{H}(2\tilde{H}+f_{1})}}
\end{displaymath}
\begin{equation}
\left. +{\frac {f_6f\tilde{v}_{R}}{f_{5}\alpha \tilde{c}_{\rm
s}r^3\tilde{H}}}(\tilde{j}-\tilde{j}_{\rm in})^2
\right\},~~~~~~~~~~~~~~~~~~~~~~~~~~~~~\label{radial_3}
\end{equation}
where
\begin{equation}
\tilde{v}_{R,\rm s}=\left\{1-{\frac
{f_1}{f_3f_4(2\tilde{H}+f_1)}}-{\frac {f_6}{f_5}}\left[1-{\frac
{f_1}{f_3f_4(2\tilde{H}+f_1)}}  \right] \right\}^{1/2}c_{\rm s},
\label{vrs_3}
\end{equation}
and
\begin{equation}
\tilde{\Omega}_{\rm K}={\frac
{1}{\sqrt{2}r^{1/2}(r-1)}};~~~~~~~~~~{\frac {1}{\tilde{\Omega_{\rm
K}}}}{\frac {d\tilde{\Omega}_{\rm K}}{dr}}=-{\frac {3r-1}{2r(r-1)}}.
\label{omega_k_d}
\end{equation}}

{The angular momentum equation (\ref{azimuthal_2}) is re-written as
\begin{equation}
\tilde{j}={\frac {\tilde{v}_R\tilde{j}_{\rm in}}{\alpha
\tilde{c}_{\rm s}\tilde{H}+\tilde{v}_{R}}}. \label{azimuthal_3}
\end{equation}
The energy equation in dimensionless form is
\begin{displaymath}
{\frac {1}{\tilde{c}_{\rm s}}}{\frac {d\tilde{c}_{\rm
s}}{dr}}={\frac {1}{f_{4}f_{5}}}\left[{\frac {2\tilde{c}_{\rm
s}^2}{f_{3}r^2\tilde{H}\tilde{\Omega}_{\rm
K}^2(2\tilde{H}+f_{1})}}-{\frac {2f_{1}}{f_{3}(2\tilde{H}+f_{1})}}
\right]{\frac {1}{\tilde{\Omega}_{\rm K}}}{\frac
{d\tilde{\Omega}_{\rm K}}{dr}}
\end{displaymath}
\begin{displaymath}
-\left[1-{\frac {f_{1}}{f_{3}f_{4}(2\tilde{H}+f_{1})}}\right]{\frac
{1} {f_{5}\tilde{v}_{R}}}{\frac {d\tilde{v}_R}{dr}}+{\frac
{2\tilde{c}_{\rm
s}^2}{f_{3}f_{4}f_{5}r^3\tilde{H}\tilde{\Omega}_{\rm
K}^2(2\tilde{H}+f_1)}}
\end{displaymath}
\begin{equation}
+{\frac
{(f_{2}-f_{3}\tilde{H}-f_{3}f_{4}\tilde{H})(2\tilde{H}+f_{1})-f_{1}\tilde{H}}{f_{3}f_{4}f_{5}r\tilde{H}(2\tilde{H}+f_{1})}}+{\frac
{f\tilde{v}_{R}}{f_{5}\alpha \tilde{c}_{\rm
s}^3r^3\tilde{H}}}(\tilde{j}-\tilde{j}_{\rm in})^2. \label{energy_5}
\end{equation}
The functions $f_1-f_6$ in the dimensionless form are given by
\begin{equation}
f_1={\frac {1}{1-{\rm e}^{-1/2}}}{\frac  {\tilde{c}_{\rm s,out}^2
\tilde{v}_R\tilde{B}_{R}^{\rm S}\tilde{B}_{z}^2\tilde{H}}
{\tilde{H}_{\rm out}r_{\rm out}^2\tilde{v}_{R,\rm
out}\tilde{\Omega}_{\rm K}^2\beta_0}}\left({\frac
{1}{\tilde{H}\kappa_0^2}}+{\frac {\xi_{\rm bz}}{\kappa_0}}\right),
\label{f_1d}
\end{equation}
\begin{displaymath}
f_{2}=-{\frac {1}{1-{\rm e}^{-1/2}}} {\frac
{\tilde{H}^2\tilde{c}_{\rm s,out}^2\tilde{v}_R}{\tilde{H}_{\rm
out}\tilde{v}_{R,\rm out}r_{\rm out}^2\tilde{\Omega}_{\rm
K}^2\beta_0(2\tilde{H}+f_1)}}~~~~~~~~~~~~~~~~~
\end{displaymath}
\begin{equation}
\times\left({\frac {2\xi_{\rm br}{\tilde{B}}_{R}^{\rm
S}\tilde{B}_z}{\tilde{H\kappa_0}}}+\xi_{\rm bz}^2\tilde{B}_R^{\rm
S}\tilde{B}_z +\xi_{\rm bz}\xi_{\rm br}\tilde{B}_R^{\rm
S}\tilde{B}_z \right), \label{f_2d}
\end{equation}
\begin{equation}
f_{3}=1-{\frac {1}{1-{\rm e}^{-1/2}}}{\frac  {\tilde{c}_{\rm
s,out}^2\tilde{v}_{R}{\tilde{B}_{R}^{\rm S}}\tilde{B}_z}
{\tilde{H}_{\rm out}\tilde{v}_{R,\rm out}r_{\rm
out}^2\tilde{\Omega}_{\rm
K}^{2}\kappa_0\beta_{0}(2\tilde{H}+f_1)}},\label{f_3d}
\end{equation}
\begin{equation}
f_{4}=1+{\frac {f_{1}}{f_{3}(2\tilde{H}+f_{1})}},\label{f_4d}
\end{equation}
\begin{equation}
f_{5}={\frac {2}{\gamma-1}}+{\frac {2\tilde{c}_{\rm
s}^2}{f_{3}f_{4}r^2\tilde{H}\tilde{\Omega}_{\rm
K}^2(2\tilde{H}+f_{1})}},\label{f_5d}
\end{equation}
and
\begin{equation}
f_6={\frac {2\tilde{c}_{\rm
s}^2}{f_3f_4r^2\tilde{H}\tilde{\Omega}_{\rm
K}^2(2\tilde{H}+f_1)}}-2.\label{f_6d}
\end{equation}
The dimensionless scale height of the disk is
\begin{equation}
\tilde{H}={\frac {1}{2}}\left ({\frac {4\tilde{c}_{\rm
s}^2}{r^2\tilde{\Omega}_{\rm K}^2}}+{f_1}^2\right)^{1/2}-{\frac
{1}{2}}f_1. \label{h_2}
\end{equation}}

\subsection{Magnetic field configuration of the ADAF}

The advection/diffusion of large-scale (poloidal) magnetic fields is
described by
\begin{equation}
{\frac {\partial}{\partial t}}[R\psi(R,0)]=-v_{R}{\frac
{\partial}{\partial R}}[R\psi(R,0)]-{\frac
{4\pi\eta}{c}}{R\over{2H}}\int\limits_{-H}^{H} J_{\phi}(R,z_{\rm
h})dz_{\rm h}, \label{indu_1}
\end{equation}
where {$J_{\phi}(R,z_{\rm h})$ is the current density at $z=z_{\rm
h}$ above/below the mid-plane of the disk, $\psi(R,z)$ is the
azimuthal component of the magnetic potential}, and the vertical
velocity $v_z$ of the flow is neglected. Assuming the azimuthal
current distribution in the vertical direction to be homogeneous, we
have
\begin{equation}
J_{\phi}(R,z_{\rm h})={\frac {J_{\phi}^{\rm s}(R)}{2H}},~~~~~
\label{js}
\end{equation}
where $J_{\phi}^{\rm s}(R)$ is the surface current density at $R$ in
the disk. As the work by \citet{1994MNRAS.267..235L}, the magnetic
field potential $\psi(R,z)=\psi_{\rm d}(R,z)+\psi_{\infty}(R,z)$,
where $\psi_{\rm d}(R,z)$ is contributed by the currents in the
accretion flow, and $\psi_{\infty}(R)=B_0R/2$ is the external
imposed homogeneous vertical field, which can be regarded as being
contributed by the currents at infinity \citep*[see][for the
details]{1994MNRAS.267..235L}. The potential $\psi_{\rm d}$ is
related to $J_{\phi}^{\rm s}(R)$ with
\begin{displaymath}
\psi_{\rm d}(R,z)={1\over c}\int\limits_{R_{\rm in}}^{R_{\rm
out}}R^\prime{\rm
d}R^\prime\int\limits_{0}^{2\pi}\cos\phi^\prime{\rm
d}\phi^\prime~~~~~~~~~~~~~~~~~~~~~~~~~~~~~~~
\end{displaymath}
\begin{displaymath}
\times\int\limits_{-H}^{H}
 {\frac {J_\phi(R^\prime,z_{\rm h})}{[{R^\prime}^2+R^2+(z-z_{\rm
h})^2-2R{R^\prime}\cos{\phi^\prime}]^{1/2}}}{\rm d}z_{\rm h}.
\end{displaymath}
\begin{displaymath}
={1\over {2Hc}}\int\limits_{R_{\rm in}}^{R_{\rm out}}J_\phi^{\rm
s}(R^\prime)R^\prime{\rm
d}R^\prime\int\limits_{0}^{2\pi}\cos\phi^\prime{\rm
d}\phi^\prime~~~~~~~~~~~~~~~~~~~~~~~~~~~
\end{displaymath}
\begin{equation}
\times\int\limits_{-H}^{H}
 {\frac {1}{[{R^\prime}^2+R^2+(z-z_{\rm
h})^2-2R{R^\prime}\cos{\phi^\prime}]^{1/2}}}{\rm d}z_{\rm h}.
\label{psi}
\end{equation}
Differentiate Eq. (\ref{psi}), we have
\begin{displaymath}
{\frac {\partial}{\partial R}}[R\psi_{\rm d}(R,z)]={1\over
{2Hc}}\int\limits_{R_{\rm in}}^{R_{\rm out}}J_\phi^{\rm
s}(R^\prime)R^\prime{\rm
d}R^\prime\int\limits_{0}^{2\pi}\cos\phi^\prime{\rm
d}\phi^\prime~~~~~~~~~~~~~~~~~~~~~~~~~~~~~~~~~~~~~~~~~~~~~~~~~~~
\end{displaymath}
\begin{equation}
\times \int\limits_{-H}^{H}
 {\frac {{R^\prime}^2+(z-{z_{\rm h}})^2-RR^\prime\cos\phi^\prime}{[{R^\prime}^2+R^2+(z-z_{\rm
h})^2-2R{R^\prime}\cos{\phi^\prime}]^{3/2}}}{\rm d}z_{\rm h}.
\label{prpsipr}
\end{equation}

For steady case, i.e., $\partial/\partial t=0$, equation
(\ref{indu_1}) becomes
\begin{equation}
-{\frac {\partial }{\partial R}}[R\psi_{\rm d}(R,0)]-{\frac
{2\pi}{c}}{\frac {\alpha c_{\rm s}R}{v_R}}{\cal P}_{\rm
m}J_{\phi}^{\rm S}(R)=B_0{R},\label{indu_2}
\end{equation}
where the magnetic Prandtl number is define as ${\cal P}_{\rm
m}=\eta/\nu$. This equation can reduce to a set of linear algebraic
equations, i.e.,
\begin{equation}
-\sum\limits_{j=1}^{n}P_{ij}{J_{\phi}^{\rm
s}(R_j)}\Delta{R_j}-{\frac {2\pi}{c}}{\frac {\alpha c_{\rm
s}(R_i)R_i}{v_R(R_i)}}{\cal P}_{\rm m}J_{\phi}^{\rm
S}(R_i)=B_0{R_i}, \label{indu_3}
\end{equation}
by using Equation (\ref{prpsipr}), where {$J_{\phi}^{\rm s}(R_{j})$
is the surface current density of the ring at radius $R_{j}$ in the
accretion flow, $\Delta{R_j}$ is the width of the ring, and}
\begin{equation}
P_{ij}= {R_j\over {2H_jc}}\int\limits_{0}^{2\pi}\cos\phi^\prime{\rm
d}\phi^\prime\int\limits_{-H_j}^{H_j}
 {\frac {{R_j}^2+{z_{\rm h}}^2-{R_i}{R_j}\cos\phi^\prime}{[{R_i}^2+{R_j}^2+z_{\rm
h}^2-2R_i{R_j}\cos{\phi^\prime}]^{3/2}}}{\rm d}z_{\rm h}.\label{pij}
\end{equation}
The surface current density $J_{\phi}^{\rm S}(R_i)$ can be
calculated by solving a set of linear algebraic equations provided
the structure of the ADAF is given with the specified magnetic
Prandtl number ${\cal P}_{\rm m}$ \citep*[see][for the
details]{1994MNRAS.267..235L}, and therefore the configuration of
the large-scale magnetic fields is available with the derived
potential $\psi$:
\begin{equation}
B_R(R,z)=-{\frac {\partial }{\partial z}}\psi(R,z),\label{b_r}
\end{equation}
and
\begin{equation}
B_z(R,z)={\frac {1}{R}}{\frac {\partial}{\partial R}}[R\psi(R,z)].
\label{b_z}
\end{equation}

\subsection{Boundary conditions}

{In order to carry out the calculation of the structure of the ADAF,
we have to specify the boundary conditions at the outer radius
$R_{\rm out}$. There are six model parameters: viscosity parameter
$\alpha$, the outer radius of the accretion disk $r_{\rm out}$, the
degree of advection $f$, specific angular momentum $\tilde{j}$ at
$r_{\rm out}$, the temperature of the gas $\Theta_{\rm out}=c_{\rm
s}/R\Omega_{\rm K}=\sqrt{2}(r-1)\tilde{c}_{\rm s}/r^{1/2}$ at
$r_{\rm out}$, and $\beta_0=8\pi p(R_{\rm out})/B_0^2$, which
relates the external imposed vertical magnetic field strength to the
gas pressure of the disk at $r_{\rm out}$. In all the calculations,
we fix the ADAF parameters: $\alpha=0.2$, $f=0.95$,
$\tilde{j}(r_{\rm out})=\tilde{j}_{\rm K}(r_{\rm out})$, and
$\Theta_{\rm out}=0.25$, which are typical for ADAFs. We find that
the final results are insensitive to the values of these parameters.
We integrate Equations (\ref{radial_3}) and (\ref{energy_5}) inwards
from $R=R_{\rm out}$ numerically. The parameter $j_{\rm in}$, the
specific angular momentum of the gas swallowed by the black hole, is
tuned carefully till the derived global solution passes the sonic
point smoothly.}

{The dynamical structure of the ADAF is coupled with the magnetic
field, i.e., the structure of the ADAF is affected by the
large-scale field threading the flow, and vice versa. In principle,
the derived solution has to pass through several critical points
(e.g., the slow/fast magneto-sonic points and the Alfv{\'e}n point).
In this work, we first calculate the structure of the ADAF without
magnetic field, and the magnetic field configuration/strength is
then calculated based on the derived ADAF structure. The structure
of the ADAF is re-calculated with this derived field
configuration/strength. The calculations of the dynamical structure
of the ADAF is decoupled with those of the field configuration, and
the global solution of the ADAF is therefore only required to pass
the sonic point. The calculations are iterated till the solutions
converge. We find that the final solution is available usually after
three or four iterations. This is because the dynamics of the ADAF
is altered little by the magnetic field except in the inner edge of
the flow.}

\section{Results}

The configuration of the large-scale magnetic fields are plotted in
Figures \ref{b_conf100_b0p1} and \ref{b_conf1000_b0p1}. The
structure of ADAFs is altered due to the presence of such
large-scale magnetic fields, which is plotted in Figures
\ref{vr_hr100}-\ref{vr_hr1000pm}. It is found that the radial
velocity of the accretion flow changes little except in the region
of the accretion flow close to the black hole horizon (see Figures
\ref{vr_hr100} and \ref{vr_hr1000}). {The accretion flow is
decelerated near the black hole by the magnetic field when $\beta_0$
is relatively small (see Figures \ref{vr_hr100}-\ref{vr_hr1000pm}),
which implies that the accretion flow can be trapped by the magnetic
field provided the external imposed field is strong enough. We find
that the accretion flow may be disrupted by the magnetic field when
the parameter $\beta_0$ is lower than a critical value $\beta_{0,\rm
crit}$, provided all other disk parameters are fixed. The critical
values of $\beta_0$ as functions of outer radius of the ADAF are
plotted in Figure \ref{beta_crit}. } The vertical structure of the
accretion flow is altered in the presence of magnetic fields, i.e.,
the scale height decreases significantly in the inner region of the
accretion flow (see the middle panels in Figures
\ref{vr_hr100}-\ref{vr_hr1000pm}). In Figures
\ref{b_kappa_r100}-\ref{b_kappa_r1000pm}, we compare the relative
importance of the gas pressure and the magnetic pressure in the
accretion flow with different magnetic field parameters (i.e.,
$\beta_0$ and ${\cal P}_{\rm m}$). {The ratios of the magnetic field
strength in the accretion flows to the external imposed field
strength as functions of radius are plotted in Figure
\ref{b_radius}. }

\vskip 1cm

\figurenum{1}
\centerline{\includegraphics[angle=0,width=8.0cm]{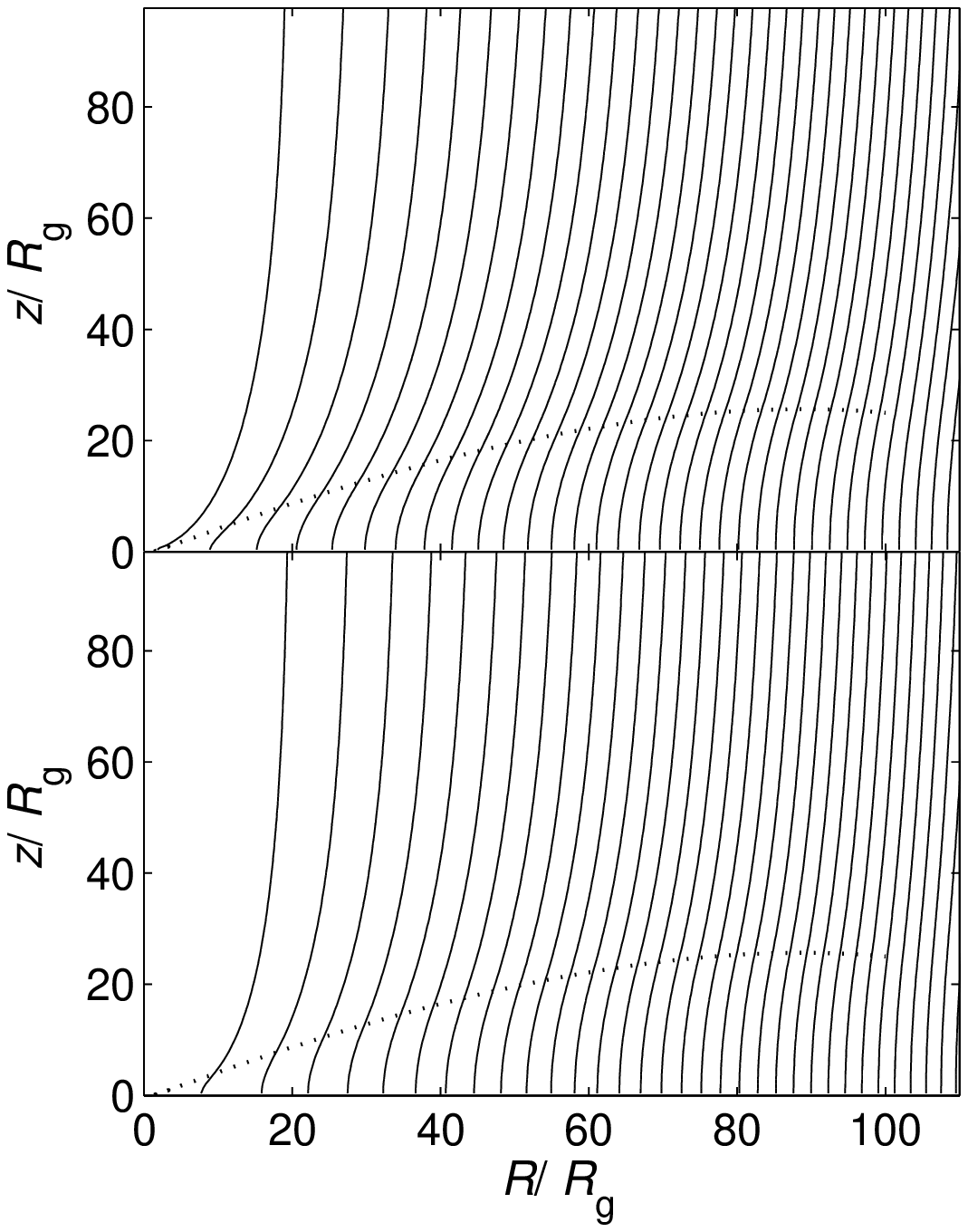}}
\figcaption{The large-scale poloidal magnetic field configuration of
an ADAF (the magnetic field strength $\beta_0=200$ is adopted). The
dotted line is the scale height of the ADAF. The outer radius of the
disk is assumed to be $R_{\rm out}=100R_{\rm g}$. The magnetic field
configuration plotted in the upper panel is calculated for the
magnetic Prandtl number ${\cal P}_{\rm m}=1$, while the lower panel
is for ${\cal P}_{\rm m}=1.5$. Every magnetic field line corresponds
to a certain value of the stream function $R\psi(R,z)$ with the
lowest $[R\psi(R,z)]_{\rm min}=0.04R_{\rm out}\psi_{\infty}(R_{\rm
out},0)$ increasing by $\Delta R\psi(R,z)=0.04R_{\rm
out}\psi_{\infty}(R_{\rm out,0})$ per line, where
$\psi_{\infty}(R_{\rm out})=B_0R_{\rm out}/2$.
\label{b_conf100_b0p1}  }\centerline{}
%\vskip 1cm
\figurenum{2}
\centerline{\includegraphics[angle=0,width=8.0cm]{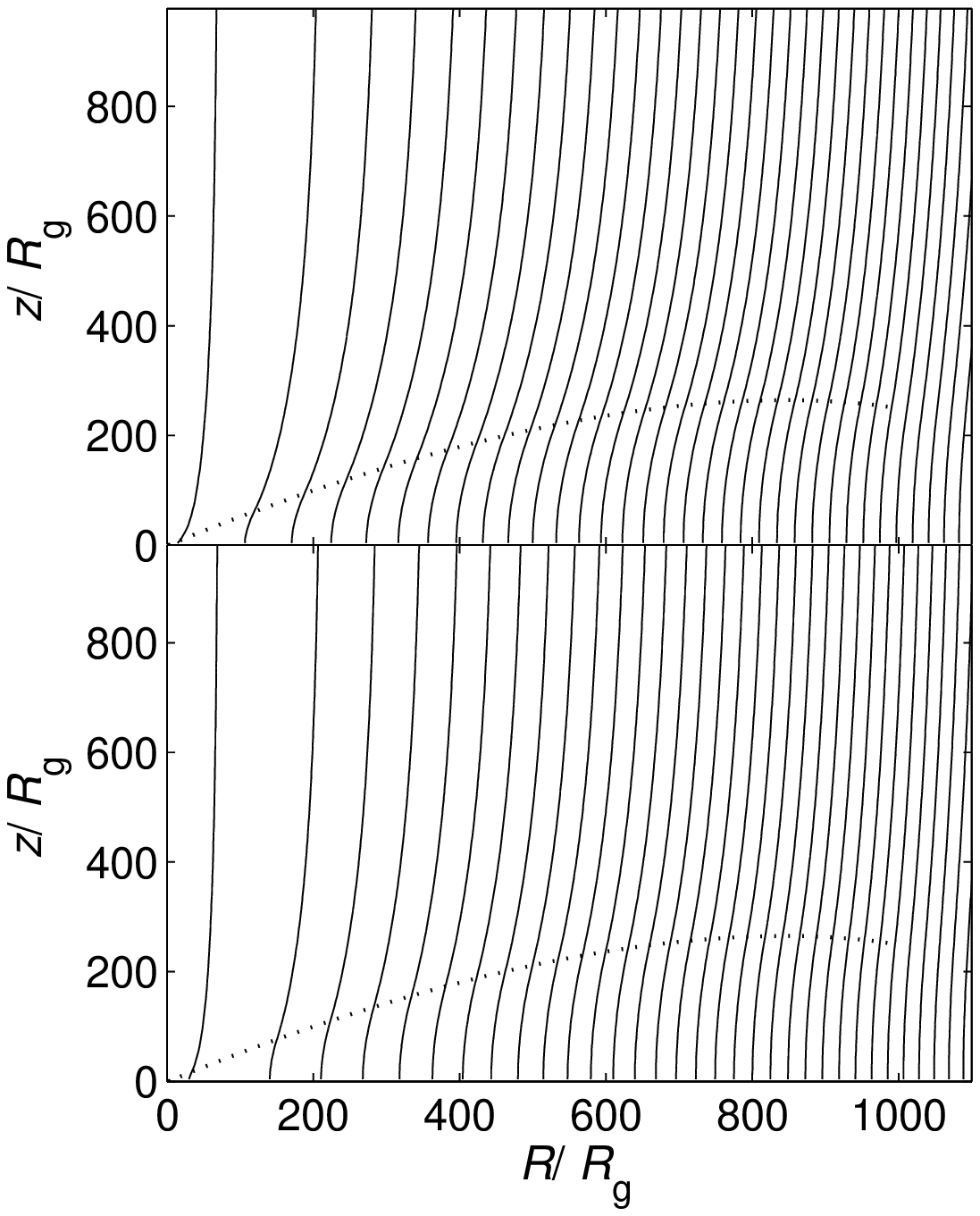}}
\figcaption{The same as Figure \ref{b_conf100_b0p1}, but the outer
radius of the disk $R_{\rm out}=1000R_{\rm g}$, and $\beta_0=20$ are
adopted. Every magnetic field line corresponds to a certain value of
the stream function $R\psi(R,z)$ with the lowest $[R\psi(R,z)]_{\rm
min}=0.005R_{\rm out}\psi_{\infty}(R_{\rm out},0)$ increasing by
$\Delta R\psi(R,z)=0.04R_{\rm out}\psi_{\infty}(R_{\rm out},0)$ per
line. \label{b_conf1000_b0p1}  }\centerline{}
%\vskip 1cm
\figurenum{3}
\centerline{\includegraphics[angle=0,width=8.0cm]{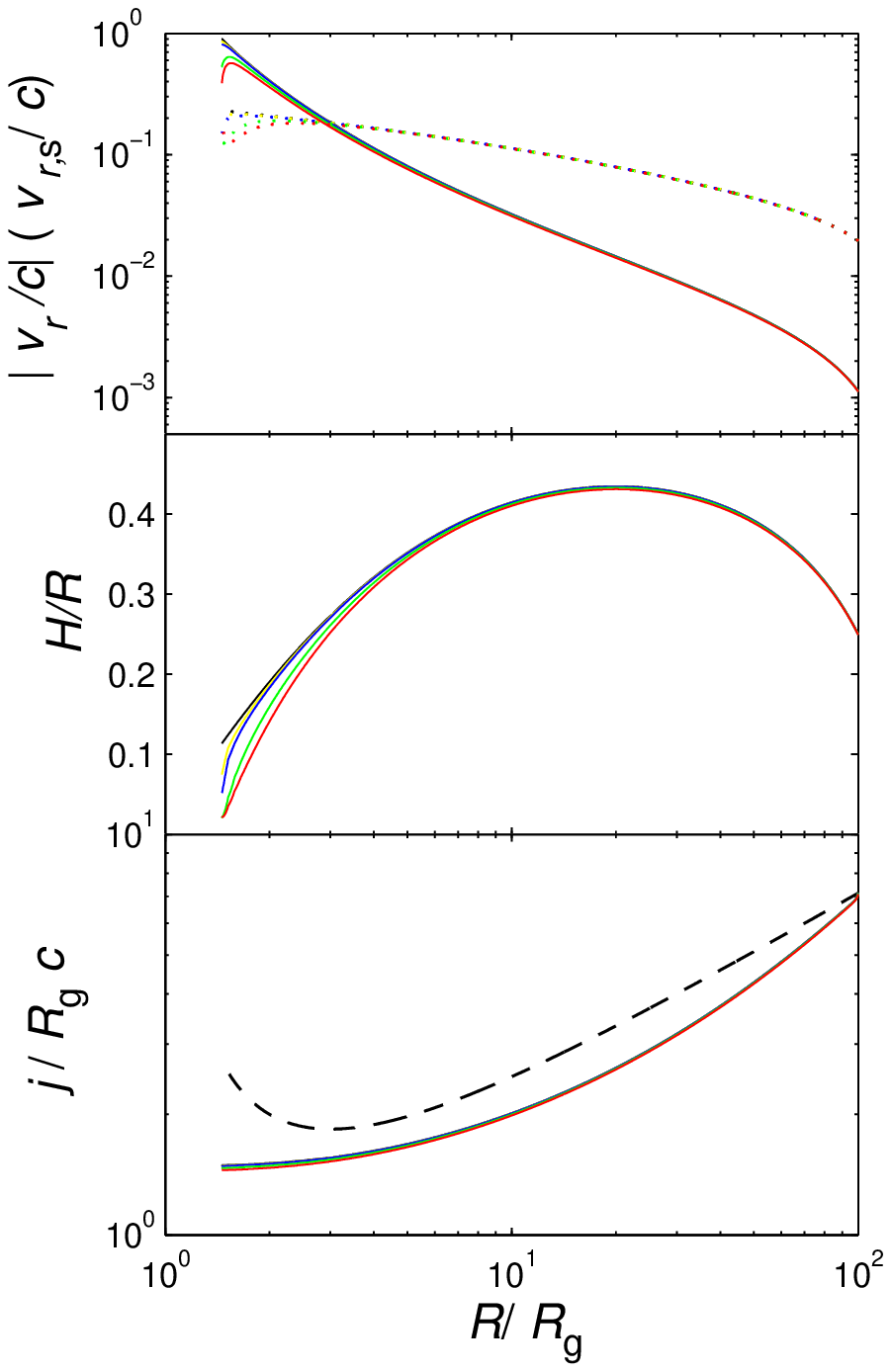}}
\figcaption{The structure of the accretion flows. The magnetic
Prandtl number ${\cal P}_{\rm m}=1$ is adopted in all the
calculations. The color lines represent the results with different
magnetic field strengths, i.e., $\beta_0=20$~(red), $40$~(green),
$200$~(blue), and $400$~(yellow). The black lines represent the
structure of the accretion flow without magnetic fields. The upper
panel: the radial velocities (solid lines), and the sound speeds
(dotted lines) as functions of radius. The middle panel: the scale
heights of accretion flows as functions of radius. The lower panel:
the specific angular momenta as functions of radius. The black
dashed line represents the Keplerian specific angular momentum.
 \label{vr_hr100}  }\centerline{}
%\vskip 1cm
\figurenum{4}
\centerline{\includegraphics[angle=0,width=8.0cm]{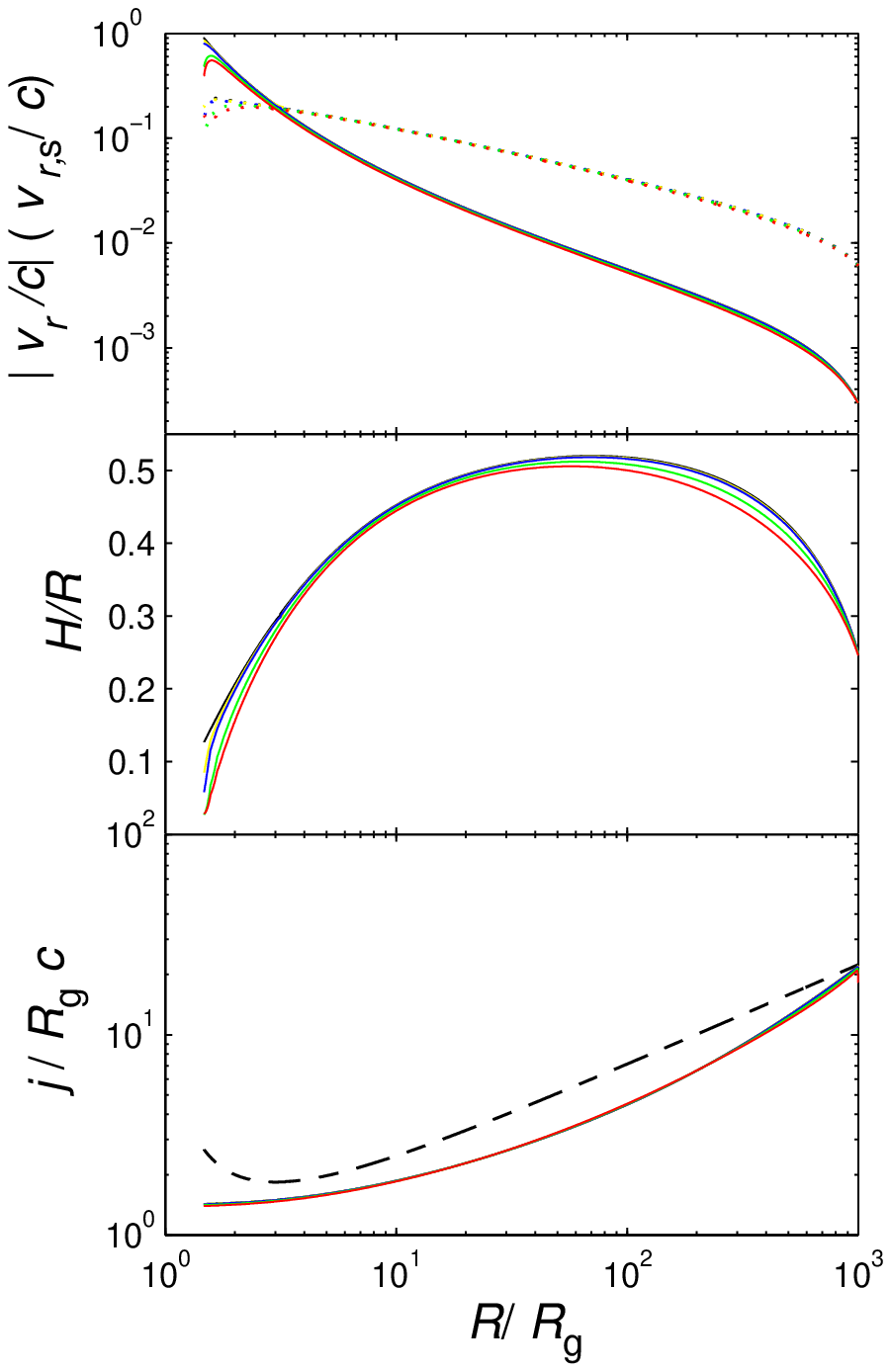}}
\figcaption{The same as Figure
 \ref{vr_hr100}, but the outer radius of the accretion flow $R_{\rm out}=1000R_{\rm g}$ is
 adopted. The color lines represent the results with different magnetic field strengths, i.e.,
$\beta_0=2$~(red), $4$~(green), $20$~(blue), and $40$~(yellow).
 \label{vr_hr1000}  }\centerline{}

\section{Discussion}

The configurations of the magnetic fields show that the advection of
magnetic fields is very efficient if ${\cal P}_{\rm m}\sim 1$ is
adopted, unlike the thin disk cases \citep{1994MNRAS.267..235L}.
This is because the radial velocity of ADAFs is much higher than
that of the thin disk. Such configurations may help launching
outflows/jets from ADAFs. For simplicity, we have not considered
magnetically driven outflows from ADAFs in this work. The radial
velocity of the ADAF will increase if the angular momentum carried
away by the outflows is properly taken into account, which will
enhance the advection of the magnetic fields in the ADAF. {We have
not considered the azimuthal component of magnetic field $B_\phi$ in
the radial momentum equation (\ref{radial_1}), which may affect the
dynamics of the accretion flow. Our results show that the dynamics
of the accretion flow is dominantly determined by the gas pressure
term except in the region very close to the black hole horizon. The
previous MHD simulations showed that the magnetic stress never
exceeds, usually much lower than, the gas pressure in the mid-plane
of the accretion flow
\citep*[e.g.,][]{2001ApJ...548..348H,2003ApJ...592.1042I}, which
implies that the main conclusions of this work will not be altered
even if the azimuthal field component is included in the radial
momentum equation. }

\vskip 1cm \figurenum{5}
\centerline{\includegraphics[angle=0,width=8.0cm]{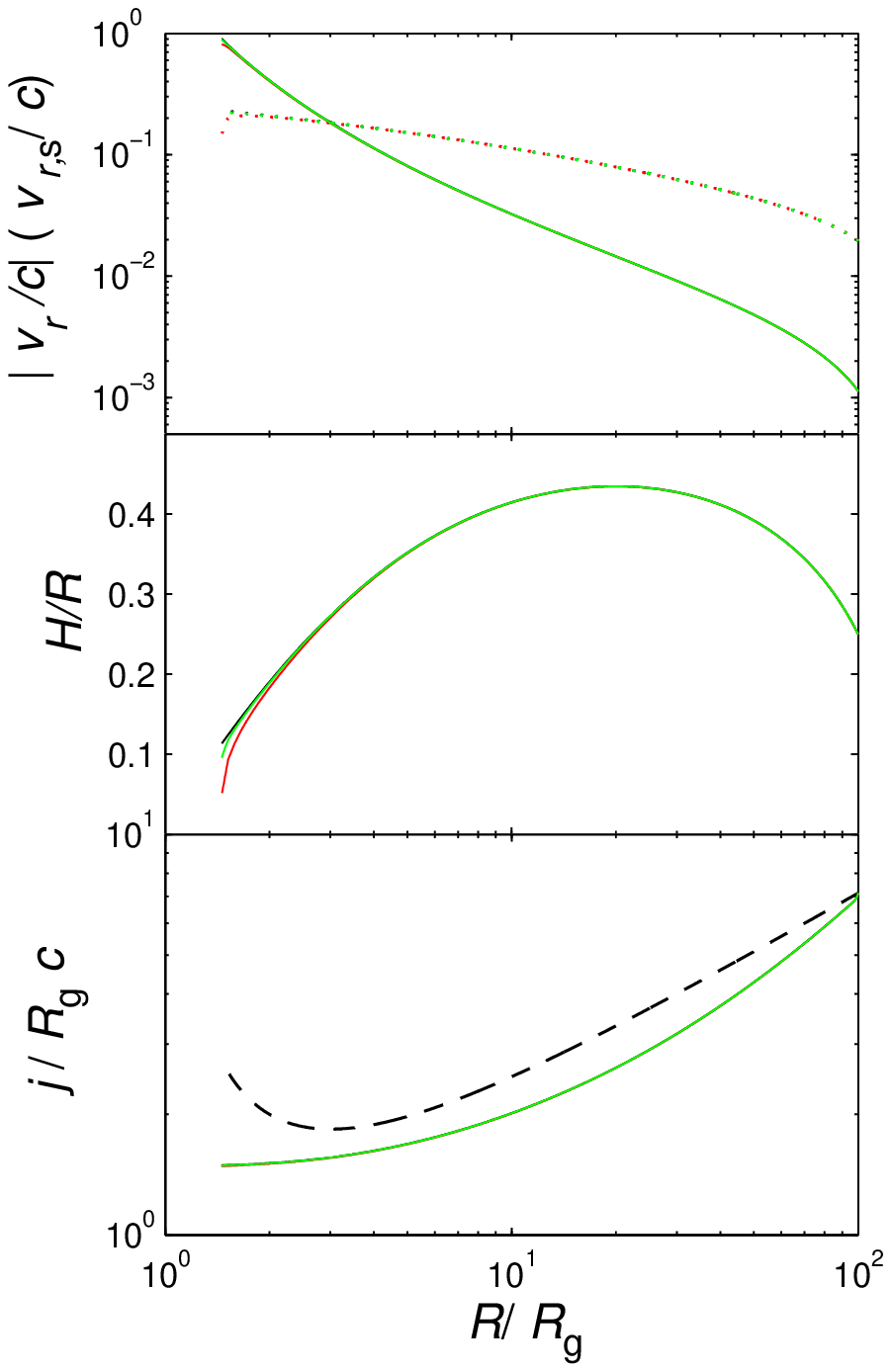}}
\figcaption{Similar to Figure
 \ref{vr_hr100}, the results with different values of the magnetic
Prandtl number are compared ($\beta_0=200$ is adopted). The red
lines represent the results calculated with ${\cal P}_{\rm m}=1$,
while the green lines are for those with ${\cal P}_{\rm m}=1.5$.
\label{vr_hr100pm}   }\centerline{}
%\vskip 1cm
\figurenum{6}
\centerline{\includegraphics[angle=0,width=8.0cm]{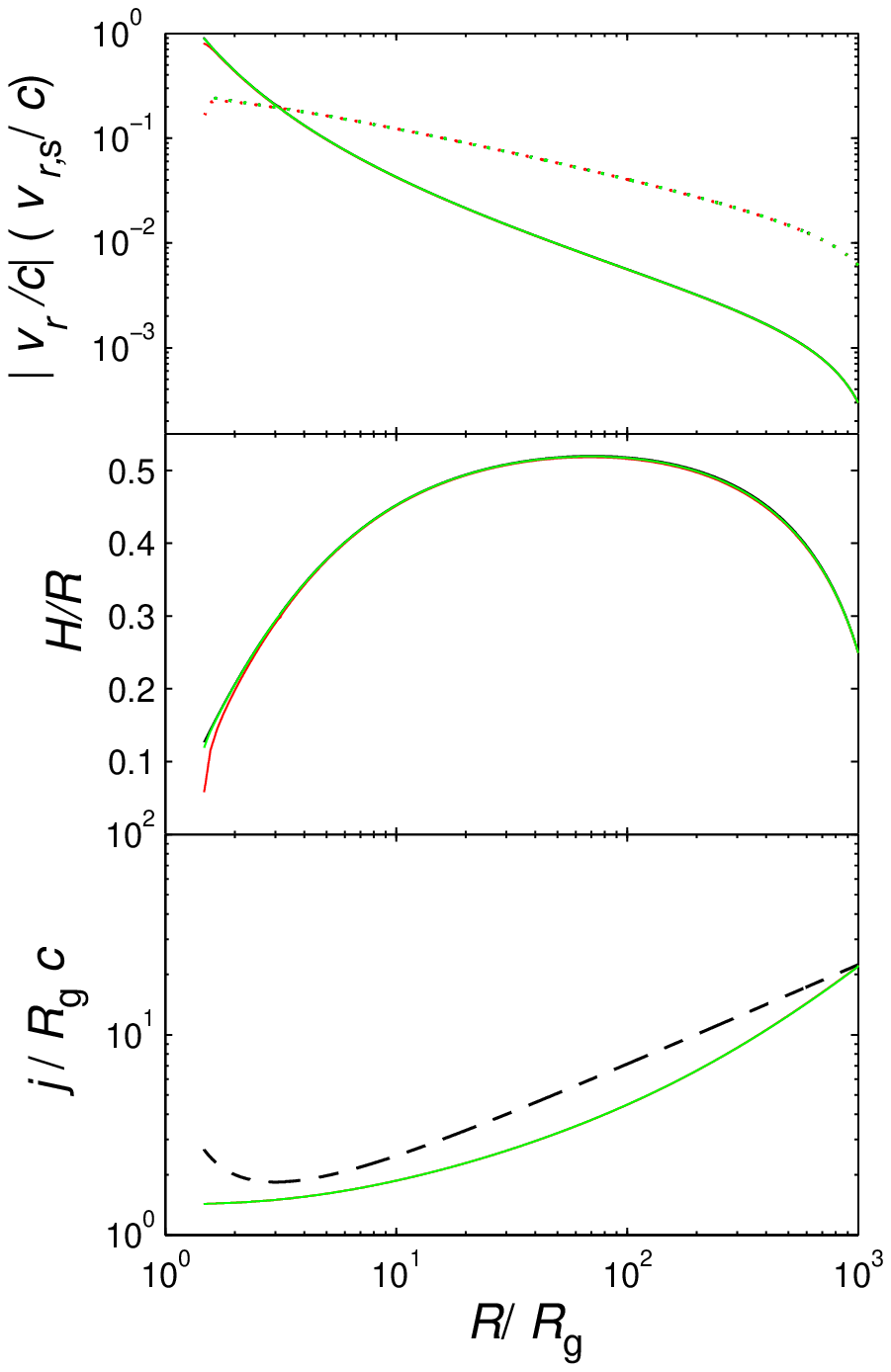}}
\figcaption{The same as Figure
 \ref{vr_hr100pm}, but the outer radius of the accretion flow $R_{\rm out}=1000R_{\rm g}$,
 and the field strength parameter $\beta_0=20$ are adopted. \label{vr_hr1000pm}   }\centerline{}
%\vskip 1cm
\figurenum{7}
\centerline{\includegraphics[angle=0,width=8.0cm]{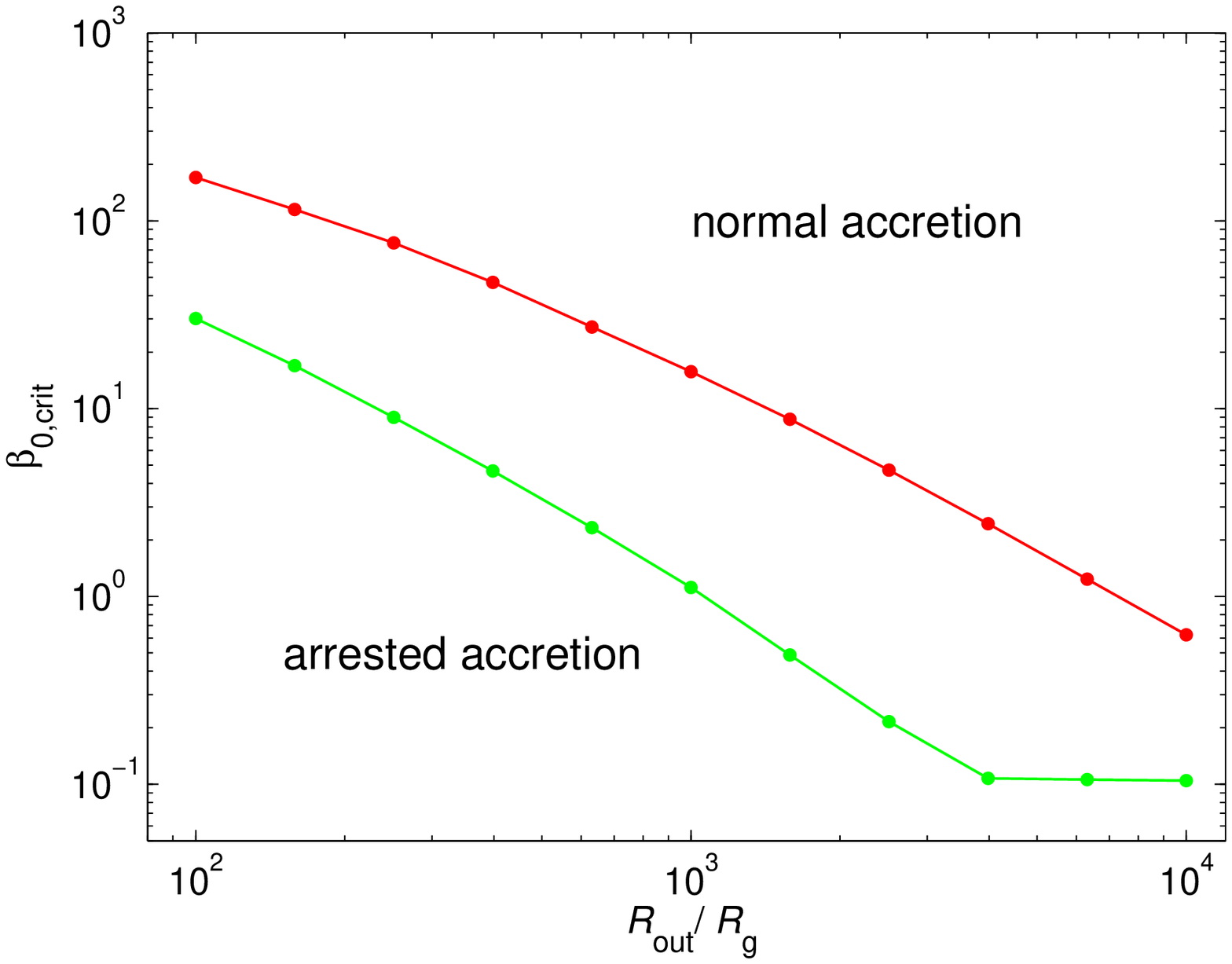}}
\figcaption{The critical values of $\beta_0$ as functions of outer
radii of ADAFs
 with ${\cal P}_{\rm m}=1$~(red) and $1.5$~(green), respectively.
\label{beta_crit}  }\centerline{}
%\vskip 1cm
\figurenum{8}
\centerline{\includegraphics[angle=0,width=8.0cm]{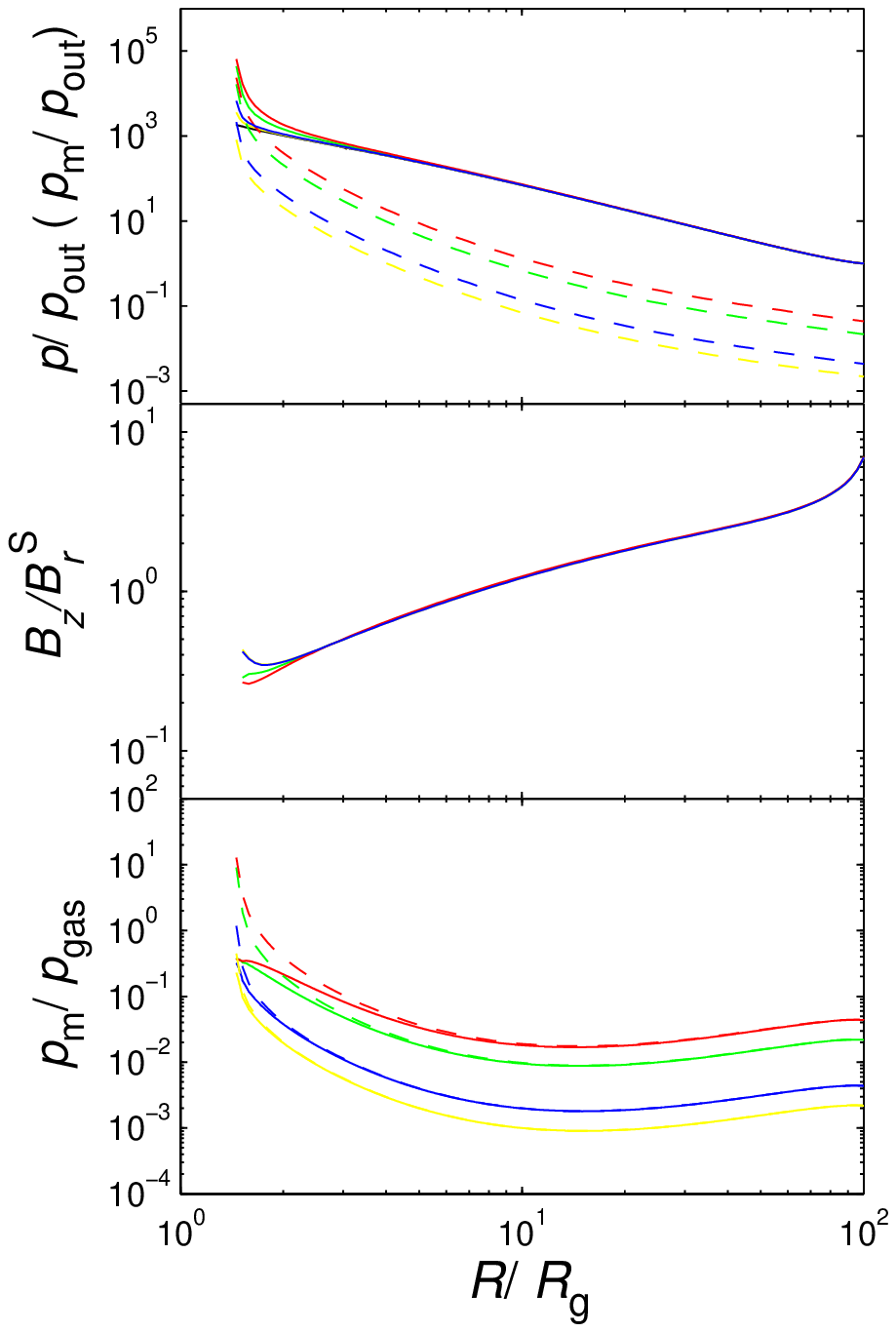}}
\figcaption{The magnetic pressure/gas pressure of the accretion
flow, and the inclination of field lines at the disk surface as
functions of radius. The magnetic Prandtl number ${\cal P}_{\rm
m}=1$ is adopted in all the calculations. The different color lines
represent the results with different magnetic field strength, i.e.,
$\beta_0=20$~(red), $40$~(green), $200$~(blue), and $400$~(yellow).
The upper panel: the magnetic field pressure/gas pressure of the
accretion flow as functions of radius. The black line represents the
gas pressure of the ADAF without magnetic fields as a function of
radius. The middle panel: the inclination of the magnetic field
lines at the disk scale height. The lower panel: the ratio of
magnetic pressure to the gas pressure in the accretion flow (solid
lines), and the ratio of magnetic pressure to the gas pressure in
the accretion flow without magnetic fields (dashed lines).
\label{b_kappa_r100}  }\centerline{}

The ratio of the magnetic pressure to gas pressure can reach $\sim
0.5$ in the inner edge of the ADAF (see Figures
\ref{b_kappa_r100}-\ref{b_kappa_r1000pm}). The radial velocity and
the temperature distribution have been altered little in the
presence of large scale magnetic fields, except in the inner region
of the ADAF ($\la 3R_{\rm g}$). However, the ADAF is vertically
pressured by the magnetic fields, which leads to the gas pressure
significantly increased in the inner region of the ADAF. It is found
that the magnetic pressure of the ADAF in the region close to the
black hole horizon can be two orders of magnitude higher than the
gas pressure of the ADAF without magnetic fields (see the dashed
lines in the lower panels in Figures \ref{b_kappa_r100} and
\ref{b_kappa_r1000}). In most previous works, the magnetic field
strength is estimated with the gas/radiation pressure in the
accretion disk on the so-called equipartition assumption, and the
confinement of the disk by the magnetic fields in the vertical
direction of the flow has not been considered
\citep*[e.g.,][]{1996MNRAS.283..854M,1997MNRAS.292..887G,1999ApJ...512..100L,1999ApJ...523L...7A,2007MNRAS.377.1652N,2008ApJ...687..156W,2009ApJ...698..594M}.
It means that the maximal jet power estimated based on the normal
accretion disk models without considering magnetically confinement
in vertical direction is around two orders of magnitude
underestimated. The assumption that the strength of the magnetic
fields near the black hole horizon should not differ significantly
in strength with the accretion disk may be incorrect, at least for
ADAFs \citep*[e.g.][]{1999ApJ...512..100L}. Our present calculations
are based on the pseudo-Newtonian potential, which can only simulate
the gravitational potential of non-rotating black holes. However,
the gas in the accretion flow plunges rapidly to the black hole in
the region of the flow with $R\la R_{\rm ms}$ ($R_{\rm ms}$ is the
radius of the marginal stable circular orbit of the hole) for a
rotating black hole, which is similar to the cases calculated in
this work for non-rotating black holes. We believe the conclusion
that the magnetic fields are significantly strengthened near the
black hole horizon due to very large radial velocity of the gas in
the flow should still hold even for rapidly spinning black holes.
This implies that the efficiency of the Blandford-Znajek mechanism
is significantly underestimated in most of the previous works
\citep*[e.g.,][]{1996MNRAS.283..854M,1997MNRAS.292..887G,1999ApJ...512..100L,2007MNRAS.377.1652N,2008ApJ...687..156W,2009ApJ...698..594M}.
The detailed calculations for rotating black holes is beyond the
scope of this work, which will be reported in our future work.

\vskip 1cm

\figurenum{9}
\centerline{\includegraphics[angle=0,width=8.0cm]{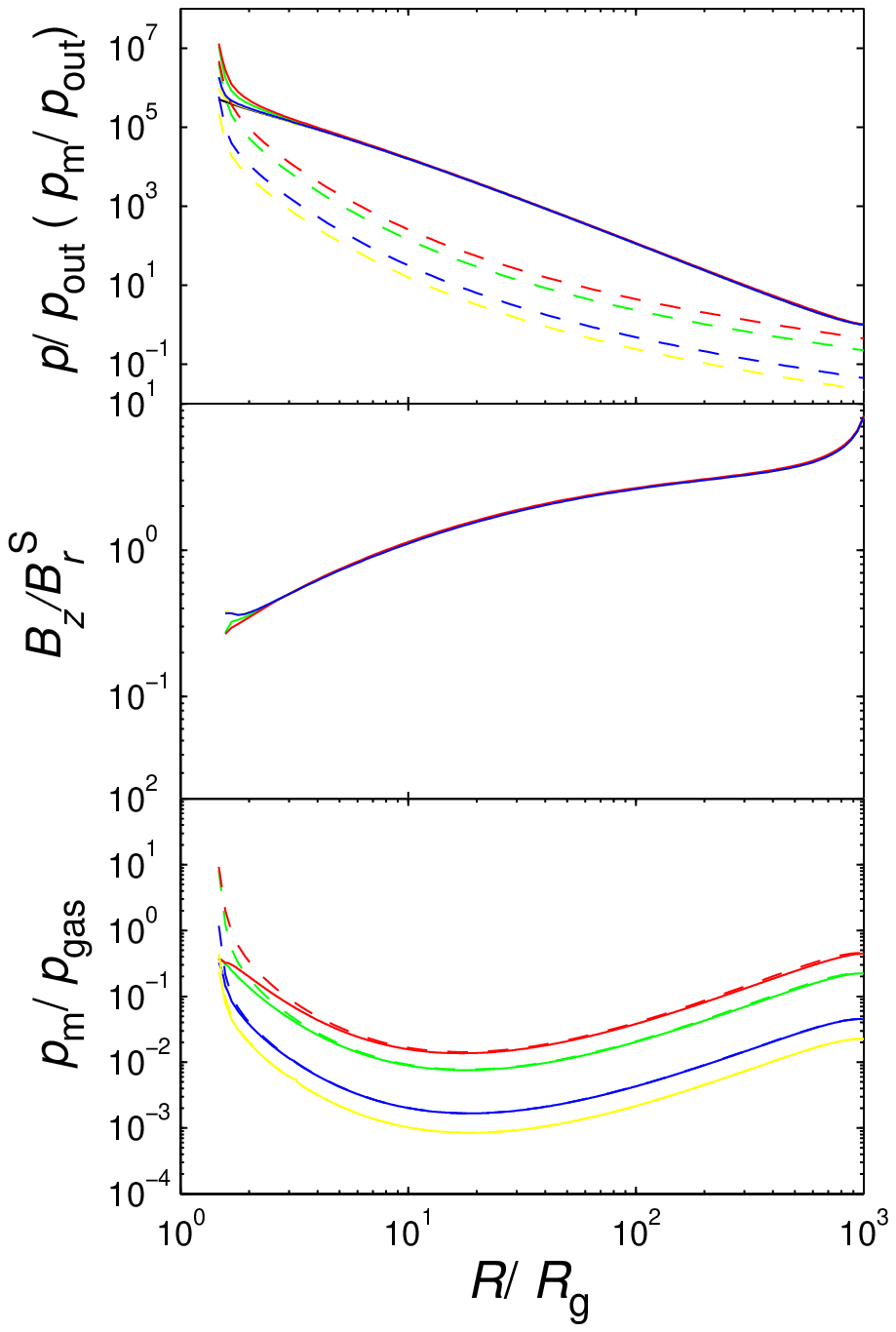}}
\figcaption{The same as Figure
 \ref{b_kappa_r100}, but the outer radius of the accretion flow $R_{\rm out}=1000R_{\rm g}$ is adopted.
 The different color lines represent the results with
different magnetic field strength, i.e., $\beta_0=2$~(red),
$4$~(green), $20$~(blue), and $40$~(yellow). \label{b_kappa_r1000}
}\centerline{}
%\vskip 1cm
\figurenum{10}
\centerline{\includegraphics[angle=0,width=8.0cm]{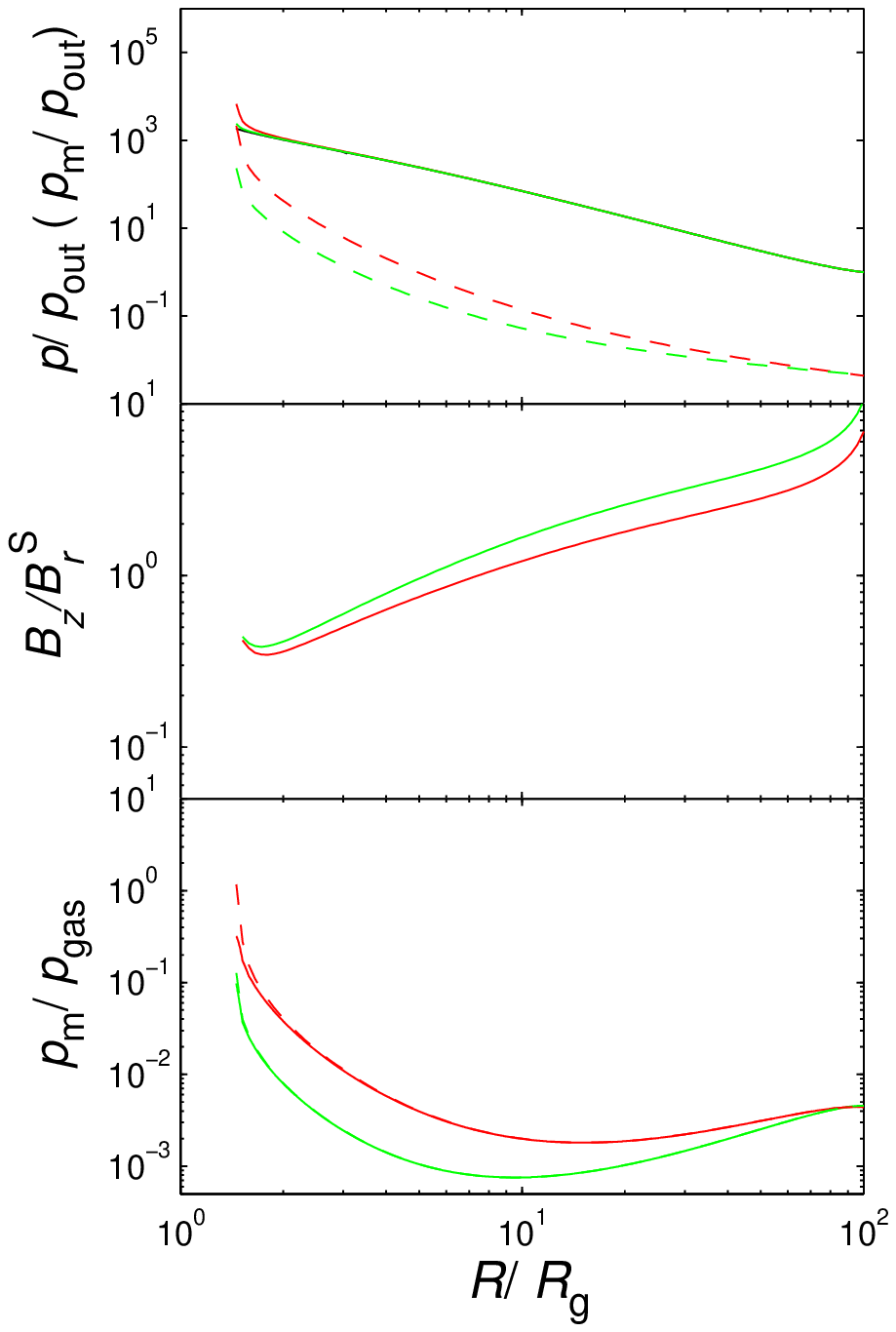}}
\figcaption{Similar to Figure
 \ref{b_kappa_r100}, the results with different values of the magnetic
Prandtl number are compared ($\beta_0=200$ is adopted). The red
lines represent the results calculated with ${\cal P}_{\rm m}=1$,
while the green lines are for those with ${\cal P}_{\rm m}=1.5$.
\label{b_kappa_r100pm}  }\centerline{}

The scale height of the ADAF is sensitive to the external imposed
homogeneous vertical magnetic field strength. We find that the scale
height of the ADAF decreases with increasing field strength, i.e.,
low-$\beta_0$ case (see the middle panels in Figures \ref{vr_hr100}
and \ref{vr_hr1000}). The scale height also depends on the value of
the magnetic Prandtl number ${\cal P}_{\rm m}$. The scale height
increases with the value of ${\cal P}_{\rm m}$ (see Figures
\ref{vr_hr100pm} and \ref{vr_hr1000pm}), because magnetic fields
being less stronger for a higher ${\cal P}_{\rm m}$ (see Figures
\ref{b_kappa_r100pm} and \ref{b_kappa_r1000pm}). The relative
thickness of the disk $H/R$ becomes very small (i.e., geometrically
thin) in the inner edge of the ADAF, due to strong magnetic
pressure. It implies that, the optical depth in the vertical
direction will increase significantly, and the temperature
difference between ions and electrons will decrease in the inner
edge of the ADAF, though no detailed radiative processes are
included in our present calculations. Its implications on the
observational features of the objects accreting at low rates can be
explored, if the detailed radiative processes are considered and the
black hole mass and the mass accretion rate need to be specified,
which is beyond the scope of this work.

\vskip 1cm \figurenum{11}
\centerline{\includegraphics[angle=0,width=8.0cm]{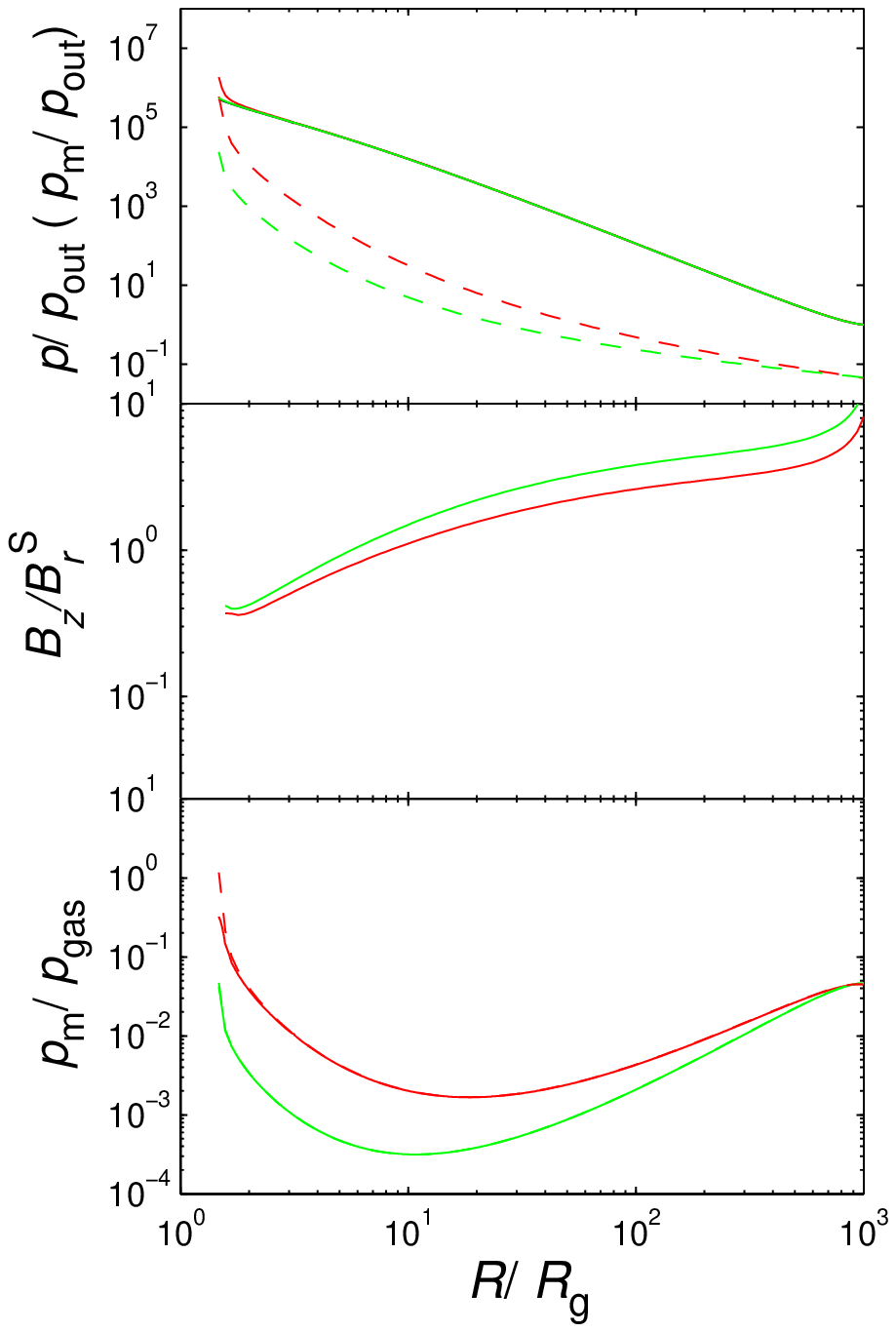}}
\figcaption{The same as Figure
 \ref{b_kappa_r100pm}, but the outer radius of the accretion flow $R_{\rm out}=1000R_{\rm g}$,
 and the magnetic field strength parameter $\beta_0=20$ are adopted.
\label{b_kappa_r1000pm}  }\centerline{}
%\vskip 1cm
\figurenum{12}
\centerline{\includegraphics[angle=0,width=8.0cm]{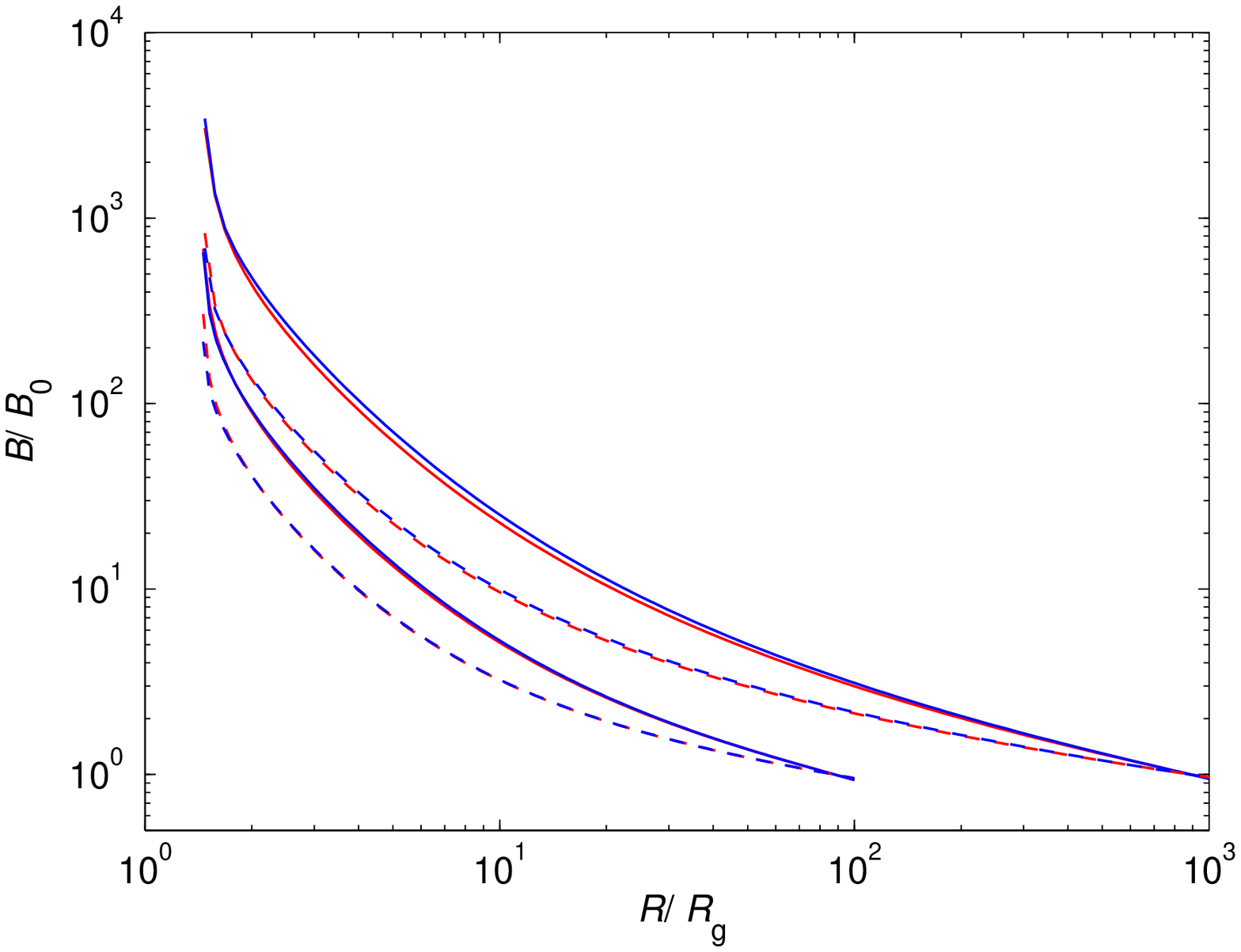}}
\figcaption{The ratios of the magnetic field strength in the
accretion
 flows to the external imposed field strength as functions of radius. The red lines represent the results with
 $\beta_0=20$ for the accretion flows with $R_{\rm out}=100R_{\rm g}$ ($\beta_0=2$ for the accretion
 flows with $R_{\rm out}=1000R_{\rm g}$), while the blue lines represent the results with $\beta_0=200$ for the
 accretion flows with $R_{\rm out}=100R_{\rm g}$ ($\beta_0=20$ for the accretion
 flows with $R_{\rm out}=1000R_{\rm g}$)  The solid lines represent the results for
 ${\cal P}_{\rm m}=1$, while the dashed lines are for ${\cal P}_{\rm m}=1.5$. \label{b_radius}  }\centerline{}

{It is interesting to find that the accretion flow is decelerated
near the black hole by the magnetic field when $\beta_0$ is
relatively small (see Figures \ref{vr_hr100} and \ref{vr_hr1000}).
There is a critical value $\beta_{0,\rm crit}$, below which the
accretion flow is decelerated near the black hole. The value of
$\beta_{0,\rm crit}$ decreases with increasing outer radius $R_{\rm
out}$, and it decreases with increasing the value of the magnetic
Prandtl number ${\cal P}_{\rm m}$ (see Figure \ref{beta_crit}). This
implies that the accretion flow may be disrupted by the magnetic
field in the inner region of the accretion flow, when the external
imposed field is strong enough or the gas pressure of the flow is
low at the outer radius, or both, which justifies the main
assumption in the qualitative analysis on the magnetic arrested
accretion disks \citep{2003PASJ...55L..69N}. In this case, the gas
may accrete as magnetically confined blobs diffusing through field
lines in the region very close to the black hole horizon
\citep*[see][for the detailed discussion, and the references
therein]{2003PASJ...55L..69N}, which is similar to those in compact
stars \citep*[e.g.,][]{1984ApJ...278..326E,1992ApJ...386...83K}. For
an ADAF surrounding a black hole with a given external imposed
homogeneous field, the value of $\beta_{0,\rm crit}$ corresponds to
a certain accretion rate at $R_{\rm out}$, below which the accretion
flow will be trapped by the magnetic field, for given temperature of
the gas in the flow at $R_{\rm out}$. Our calculations show that the
gas can alternatively be trapped by the magnetic field at outer
radius of the accretion flow provided the magnetic pressure
significantly exceeds over the gas pressure, i.e., $\beta_0\ll 1$
(see the green line in Figure \ref{beta_crit} at large radii). We
compare the amplification of the magnetic field advected in the
accretion flows with different values of the disk parameters, and
find that the amplification of the field is dominantly determined by
the value of ${\cal P}_{\rm m}$, while it is insensitive to the
adopted value of $\beta_0$. It implies that the strength of the
field near the black hole horizon is mainly determined by the
strength of the external imposed field and the outer radius of the
ADAF, which is qualitatively consistent with the results for thin
accretion disks in \citet{1994MNRAS.267..235L}, though the advection
of the field is rather inefficient in the thin disk case. This means
that the strength of the magnetic field near the black hole can be
estimated if the strength of the ordered field threading the ambient
gas and the outer radius $R_{\rm out}$ of the ADAF are known. It is
found that stronger magnetic field will be in the inner edge of the
ADAF, if the gas starts to accrete from a larger radius (see Figure
\ref{b_radius}). This should be very useful for estimate of the
field strength near the black hole horizon in some low-luminosity
active galactic nuclei, of which the properties of the ambient gas
at the Bondi radius have well been measured with observations
\citep*[e.g.,][]{2003ApJ...591..891B,2006MNRAS.372...21A}. }

{We notice that a series of works of numerical simulations on the
magnetic fields in ADAFs
\citep*[][]{2002ApJ...566..137I,2003ApJ...592.1042I,2006ApJ...649..361I,2008ApJ...677..317I}.
They found that the initial uniform $z$-direction magnetic field can
be efficiently dragged inwards by the ADAF, and their derived
magnetic field configuration is similar to those obtained in this
work \citep*[see Figure 3 in][]{2002ApJ...566..137I}. The numerical
simulations show that the magnetic field in the inner edge of the
accretion can be very strong to disrupt the accretion flow
\citep{2008ApJ...677..317I}, which agrees qualitatively with our
calculations. Compared with these MHD simulations on ADAFs, our
calculations explicitly indicate that the vertical scale-height of
the ADAF near the black hole is significantly reduced, and hence the
field strength can be much higher than that estimated based on the
conventional ADAF model.  }

In this work, the viscosity $\alpha=0.2$ is fixed. The radial
velocity of the ADAF in the outer region is roughly proportional to
the kinetic viscosity $\nu$, and the magnetic Prandtl number ${\cal
P}_{\rm m}=\eta/\nu$, which implies the advection/diffusion process
in the accretion flow is almost independent of the adopted value of
$\alpha$. The dynamical structure of ADAFs is insensitive to the
value of $f$ provided the accretion flow is advection dominated,
i.e., $0.5\la f\la 1$. We also calculate the problem with different
values of the disk parameters (e.g., $f$, $\Theta_{\rm out}$, and
$\tilde{j}_{\rm out}$), and find that the main conclusions have not
been altered.

The ADAF may connect to a thin accretion disc at a certain
transition radius $R_{\rm tr}$. This is required by modeling on a
variety of observations of AGNs/X-ray binaries
\citep*[e.g.][]{1997ApJ...489..865E,1999ApJ...525L..89Q,2000ApJ...537L.103L,2003ApJ...599..147C,2004ApJ...612..724Y,2009RAA.....9..401X},
and is also predicted by some theoretical model calculations
\citep*[e.g.,][]{1995ApJ...438L..37A,1999ApJ...527L..17L,2000A&A...360.1170R,2002A&A...387..918S,2004ApJ...602L..37L}.
Our calculations are also valid for the case that the inner ADAF
connects to the outer cold thin disk at a certain radius. In this
case, the advection of the external fields is quite inefficiently in
the outer thin disk due to its low radial velocity, and the field
lines thread the disk almost vertically \citep*[see,
e.g.,][]{1994MNRAS.267..235L}, while these field lines can be
efficiently dragged inward by the radial motion of the inner ADAF.

\acknowledgments  I thank Henk Spruit, Dong Lai, and Qingwen Wu for
helpful discussion. This work is supported by the NSFC (grants
10821302, and 10833002), the National Basic Research Program of
China (grant 2009CB824800), the Science and Technology Commission of
Shanghai Municipality (10XD1405000), and the CAS/SAFEA International
Partnership Program for Creative Research Teams.

{}

\end{document}